\documentclass[twocolumn,aps,prr,amsmath,amssymb,preprintnumbers,superscriptaddress,longbibliography]{revtex4-2}
\DeclareUnicodeCharacter{2212}{-}
\usepackage{physics}
\usepackage[normalem]{ulem}
\usepackage{braket}
\usepackage{bbold}
\usepackage{amsmath, amsfonts, amssymb}
\usepackage{graphicx}
\usepackage{float}
\usepackage{lipsum}
\usepackage[colorlinks=true,linkcolor=blue,citecolor=blue,allcolors=blue]{hyperref}
\usepackage[caption=false]{subfig}
\usepackage{mathtools}
\usepackage{tikz}
\usepackage{notes2bib}
\bibnotesetup{
	note-name = ,
	use-sort-key = false
}
\begin{document}
	\newcommand{\titleinfo}{ Ab-Initio Approach to Many-Body Quantum Spin Dynamics}
	\title{\titleinfo}

        \author{Aditya Dubey}
        \affiliation{Department of Physics, BITS-Pilani, K K Birla Goa Campus, Zuarinagar, Goa 403726, India}
        \affiliation{Zentrum f{\"u}r Optische Quantentechnologien, Universit{\"a}t Hamburg, Luruper Chaussee 149, 22761 Hamburg, Germany}
 
        \author{Zeki Zeybek}
	\email{zeki.zeybek@uni-hamburg.de}
	\affiliation{Zentrum f{\"u}r Optische Quantentechnologien, Universit{\"a}t Hamburg, Luruper Chaussee 149, 22761 Hamburg, Germany}
        \affiliation{The Hamburg Centre for Ultrafast Imaging, Universit{\"a}t Hamburg, Luruper Chaussee 149, 22761 Hamburg, Germany}

        \author{Fabian K{\"o}hler}
	\affiliation{SpeQtral, 1 Fusionopolis Way, $\#$05-02/03/04 Connexis South, Singapore 138632, Singapore}

	\author{Rick Mukherjee}
	\email{rmukherj@physnet.uni-hamburg.de}
	\affiliation{Zentrum f{\"u}r Optische Quantentechnologien, Universit{\"a}t Hamburg, Luruper Chaussee 149, 22761 Hamburg, Germany}

        \author{Peter Schmelcher}
        \email{peter.schmelcher@uni-hamburg.de}
\affiliation{Zentrum f{\"u}r Optische Quantentechnologien, Universit{\"a}t Hamburg, Luruper Chaussee 149, 22761 Hamburg, Germany}	
 \affiliation{The Hamburg Centre for Ultrafast Imaging, Universit{\"a}t Hamburg, Luruper Chaussee 149, 22761 Hamburg, Germany}
 
	\begin{abstract}
	    A fundamental longstanding problem in studying spin models is the efficient and accurate numerical simulation of the long-time behavior of larger systems. The exponential growth of the Hilbert space and the entanglement accumulation at long times pose major challenges for current methods. To address these issues, we employ the multilayer multiconfiguration time-dependent Hartree (ML-MCTDH) framework to simulate the many-body spin dynamics of the Heisenberg model in various settings, including the Ising and XYZ limits with different interaction ranges and random couplings. Benchmarks with analytical and exact numerical approaches show that ML-MCTDH accurately captures the time evolution of one- and two-body observables in both one- and two-dimensional lattices. A comparison with the discrete truncated Wigner approximation (DTWA) highlights that ML-MCTDH is particularly well-suited for handling anisotropic models and provides more reliable results for two-point observables across all tested cases. The behavior of the corresponding entanglement dynamics is analyzed to reveal the complexity of the quantum states. Our findings indicate that the rate of entanglement growth strongly depends on the interaction range and the presence of disorder. This particular relationship is then used to examine the convergence behavior of ML-MCTDH. Our results indicate that the multilayer structure of ML-MCTDH is a promising numerical framework for handling the dynamics of generic many-body spin systems.
	\end{abstract}
	\maketitle

\section{Introduction}

Many-body spin models are crucial for a variety of applications ranging from understanding the behavior of magnetic materials \cite{RevModPhys.25.353,10.1143/PTP.35.16,PhysRevLett.78.1984}, disordered systems \cite{I_Ya_Korenblit_1978,PhysRevLett.113.107204,PhysRevA.80.032311} to strongly correlated phases \cite{Savary_2017,PhysRevB.88.125122,doi:10.1126/science.abi8794}, quantum sensing \cite{RevModPhys.89.035002} and quantum information processing \cite{Bennett2000,PhysRevA.52.3457,PhysRevLett.89.147902}. Exploring the time evolution of these models provides insights into fundamental aspects of quantum dynamics \cite{Blaß2016,PhysRevE.93.032104,10.1063/1.4969869}, including the emergence of long-range correlations \cite{Lieb1972,PhysRevLett.85.3233,PhysRevA.108.023301}, entanglement \cite{PhysRevA.69.022304,PhysRevA.75.012325,PhysRevA.71.052308,PhysRevLett.92.027901}, and information spreading \cite{PhysRevA.69.052315,PhysRevLett.91.207901,PhysRevA.71.032314}. Although many-body spin systems are accessible in controlled experimental settings such as trapped ions \cite{Blatt2012,Britton2012,RevModPhys.93.025001,PhysRevA.99.052342,PhysRevA.100.059902}, Rydberg atoms \cite{Labuhn2016,PhysRevX.7.041063,PhysRevLett.128.113602}, and polar molecules \cite{Yan2013,Li2023,doi:10.1126/science.abd9547}, numerical simulation methods remain essential in accompanying experimental and theoretical investigations. For example, controlled experimental platforms can still suffer from unwanted noise \cite{RevModPhys.86.153,PhysRevX.10.041038,PhysRevA.78.052325} while numerical simulations provide useful insights into isolated systems that are free from environmental effects. Numerical simulations can reach regimes that might exceed the limits of experimental resolution.

However, computational simulation of spin dynamics poses significant challenges: $i)$ exponential scaling of the Hilbert space with system size, and $ii)$ difficulty to accurately describe long time dynamics. Exact methods such as exact diagonalization (ED) are limited to small system sizes since the Hilbert space dimension grows as $2^N$ for a spin-$1/2$ system with $N$ spins. Often it is the case that when many-body quantum systems evolve, they accumulate entanglement \cite{RevModPhys.80.517,PhysRevB.95.094302}, making numerical methods built on entanglement restriction such as matrix-product-state (MPS) \cite{SCHOLLWOCK201196,ORUS2014117} based time-dependent density matrix renormalization group (tDMRG) \cite{PhysRevLett.93.040502, PhysRevLett.93.076401,Daley_2004} and its variants \cite{PhysRevLett.107.070601,PhysRevLett.109.267203,PhysRevB.94.165116} suitable only for short-time dynamics. Semiclassical phase-space methods such as discrete truncated Wigner approximation (DTWA) \cite{PhysRevX.5.011022} have been widely applied in various contexts \cite{Schachenmayer_2015,PhysRevResearch.3.013060,PhysRevB.102.014303,PhysRevA.99.043627,Shenoy_2024}. While DTWA is effective in capturing short-time dynamics, it can struggle in reproducing certain kinds of spin dynamics, and in particular, anisotropic models.

    \begin{figure}
 \centering
\begin{tabular}{c}
  \includegraphics[width=85mm]{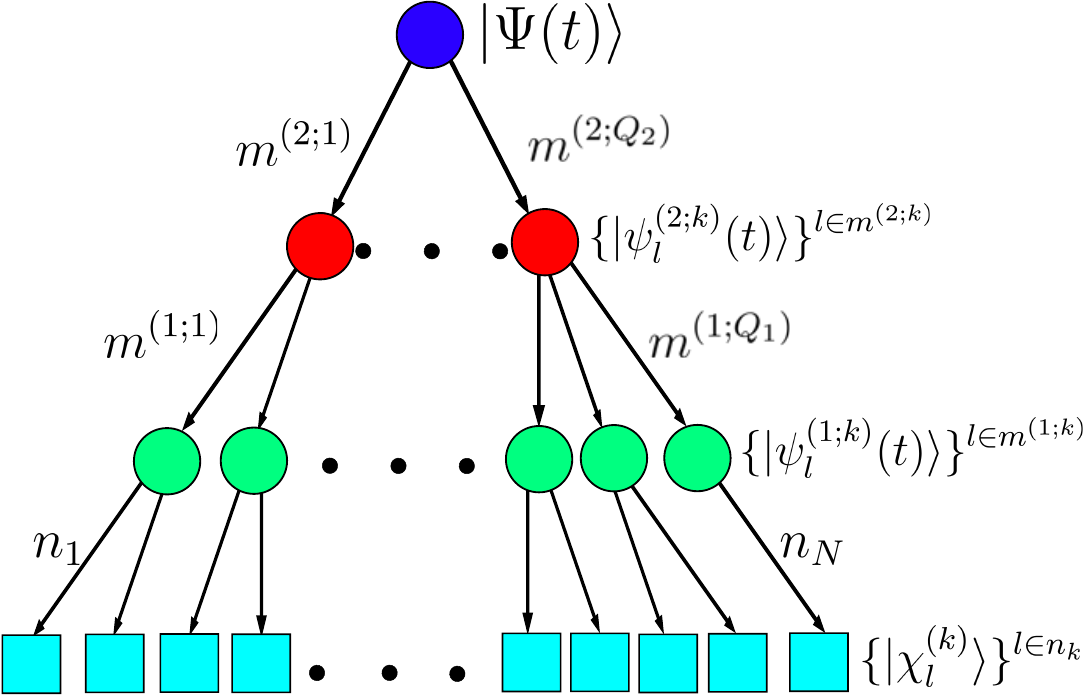}
\end{tabular}
\caption{ Diagrammatic representation of the two-layer ML-MCTDH ansatz for the many-body wavefunction $\ket{\Psi(t)}$ of a system with $N$ degrees of freedom. 
The cyan boxes represent the time-independent basis states, while the green and red circles correspond to the time-dependent basis states of layers 1 and 2, respectively. The blue circle denotes the many-body wavefunction.} 
\label{ML_Tree}
\end{figure}

To overcome some of the issues mentioned above, we propose using the ML-MCTDH \cite{10.1063/1.1580111,10.1063/1.2902982,10.1063/1.2363195,Wang2007-ds,doi:10.1021/jp072217m} method which employs a small number of time-dependent basis states to represent the many-body wavefunction, allowing for a controlled truncation of the Hilbert space. ML-MCTDH builds on the ideas of multiconfiguration time-dependent Hartree (MCTDH) \cite{MEYER199073,10.1063/1.463007,BECK20001} approach, which was initially developed for studying high-dimensional wave packet dynamics of complex molecular systems \cite{10.1063/1.472327,10.1063/1.1436307,doi:10.1126/science.1104085}. Subsequent developments made it possible to treat many-body systems comprising identical particles, including bosons \cite{PhysRevLett.99.030402,PhysRevA.77.033613,10.1063/1.3173823,10.1063/1.4975662,10.1063/1.5140984, niermann_multi-layer_2024} and fermions \cite{JürgenZanghellini_2004,PhysRevA.71.012712,10.1063/5.0028116}, as well as mixtures thereof \cite{10.1063/1.4821350,Krönke_2013,10.1063/1.4993512}. Recently, it has also been successfully employed to determine the ground-state properties of disordered spin models using the imaginary time propagation scheme \cite{PhysRevResearch.5.023135}, as well as to study the dynamics of centrally coupled systems with many-body interacting baths \cite{PhysRevB.103.134201}. 

Inspired by these promising results, we apply the ML-MCTDH framework to investigate the time evolution of the Heisenberg model in different scenarios, including the Ising and the highly anisotropic XYZ cases. In the Ising limit, comparisons with analytical results indicate that ML-MCTDH matches with exact solutions for the time evolution for all interaction ranges. In particular, it outperforms DTWA at depicting the dynamics of two-point observables in both 1D chains and 2D square lattices. In the XYZ case, comparisons with exact diagonalization (ED) results show that the ML-MCTDH captures the dynamics of both one- and two-point observables exactly for all interaction ranges while DTWA only captures short-time dynamics of one-point observables and fails to depict the two-point observable. In the disordered Ising, benchmarks with analytical results show that the ML-MCTDH becomes exact for power-law interactions with dipolar and Van der Waals characters, especially outperforming DTWA for the two-point observable. The analysis of entanglement dynamics is carried out to understand the complexity of quantum states. Our results show that the rate at which entanglement grows after a quench is heavily dependent on the interaction range and the presence of disorder. In the absence of disorder, for 
$\alpha=3$, the entanglement entropy increases linearly with time, whereas for 
$\alpha=0$, it exhibits an oscillatory behavior. In the case of Ising disorder, interactions with 
$\alpha=3$ lead to a logarithmic increase, while all-to-all interactions result in a rapid rise in entanglement before saturating near the theoretical maximum.

This paper is organized as follows. We provide a brief introduction to ML-MCTDH in Sec.~\ref{ML_MCTDH_text} and discuss the different limits of the Heisenberg model for which we evaluate the dynamics in Sec.~\ref{spin_models}. Section~\ref{Results} contains the results of our approach for the dynamics of collective one- and two-point observables, which are benchmarked against exact methods and DTWA. We also discuss the entanglement growth dynamics and discuss its relationship with the convergence behavior of ML-MCTDH. Finally, in section~\ref{Conclusion} we provide our conclusions and outlook. The appendix~\ref{MLMCTDH} details the truncation scheme employed in ML-MCTDH, and appendix~\ref{DTWA} contains a brief description of DTWA.

\section{ Computational approach and spin models } 
\label{Theory}

ML-MCTDH is a well-established numerical ab initio method previously applied to a variety of physics and chemistry problems \cite{10.1063/1.1580111,10.1063/1.2902982,10.1063/1.2363195,Wang2007-ds,doi:10.1021/jp072217m,PhysRevLett.99.030402,PhysRevA.77.033613,10.1063/1.3173823,10.1063/1.4975662,10.1063/1.5140984,JürgenZanghellini_2004,PhysRevA.71.012712,10.1063/5.0028116,10.1063/1.4821350,Krönke_2013,10.1063/1.4993512} and thus the technical details of this method are provided in Appendix \ref{MLMCTDH}. Here, we focus on the essential features of ML-MCTDH that relate to the numerical study of spin dynamics.
\subsection{Multilayer multiconfiguration
time-dependent Hartree method}
\label{ML_MCTDH_text}

The standard approach for solving the time-dependent Schrödinger equation involves propagating a wavepacket. For a spin-$1/2$ system, this wavepacket ansatz is typically expressed as $\ket{\Psi(t)} = \sum\limits_{s_1,\cdots,s_N}A_{s_1 \cdots s_N}(t)\ket{s_1,\cdots,s_N}$ with $s_i \in \{\uparrow,\downarrow\}$. $A_{s_1 \cdots s_N}$ denotes the time-dependent expansion coefficients, $\ket{s_1,\cdots,s_N}$ are time-independent basis states which exhibit exponential scaling with the system size. ML-MCTDH instead constructs an ansatz $\ket{\Psi(t)}$ (explicitly given in \ref{L2Ansatz}) in the form of a hierarchical network, resulting in a tree structure with multiple layers consisting of a set of nodes as shown in Fig.~\ref{ML_Tree}. Each node groups a certain number of spin degrees of freedom ($\chi_l^{(k)}$) into a single unit described by a small set of time-dependent basis states  with time-dependent expansion coefficients as denoted by green ($\psi_l^{(1;k)}$) and red ($\psi_l^{(2;k)}$) nodes in Fig.~\ref{ML_Tree}, thereby addressing the exponential scaling of the Hilbert space. Choosing the optimal tree structure and number of time-dependent basis states allows for a compact representation of the wavefunction that spans the subspace necessary for accurately capturing the dynamics. {This flexibility allows ML-MCTDH to represent a wide range of systems. However, constructing the optimal tree structure for arbitrary interactions and higher-dimensional systems becomes a challenging task.} The time-dependent expansion coefficients are updated using the Dirac-Frenkel variational principle \cite{Dirac_1930,BECK20001,10.1063/1.1580111}. More details about the truncation scheme are provided in Appendix \ref{MLMCTDH}.

\begin{figure*}
\begin{center}
\begin{tabular}{ccc}
  \includegraphics[width=56mm]{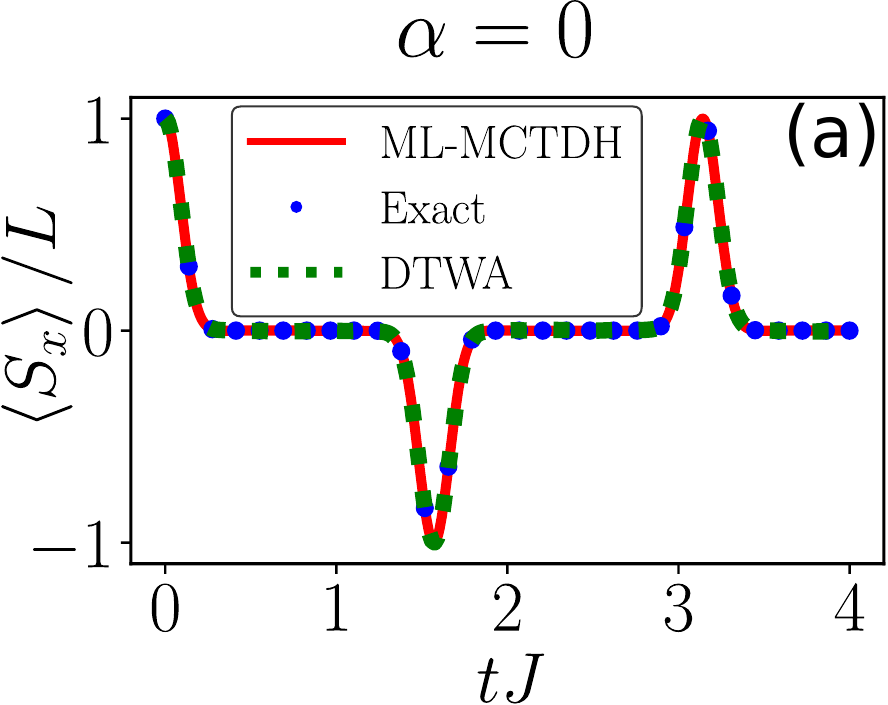}& \includegraphics[width = 55mm]{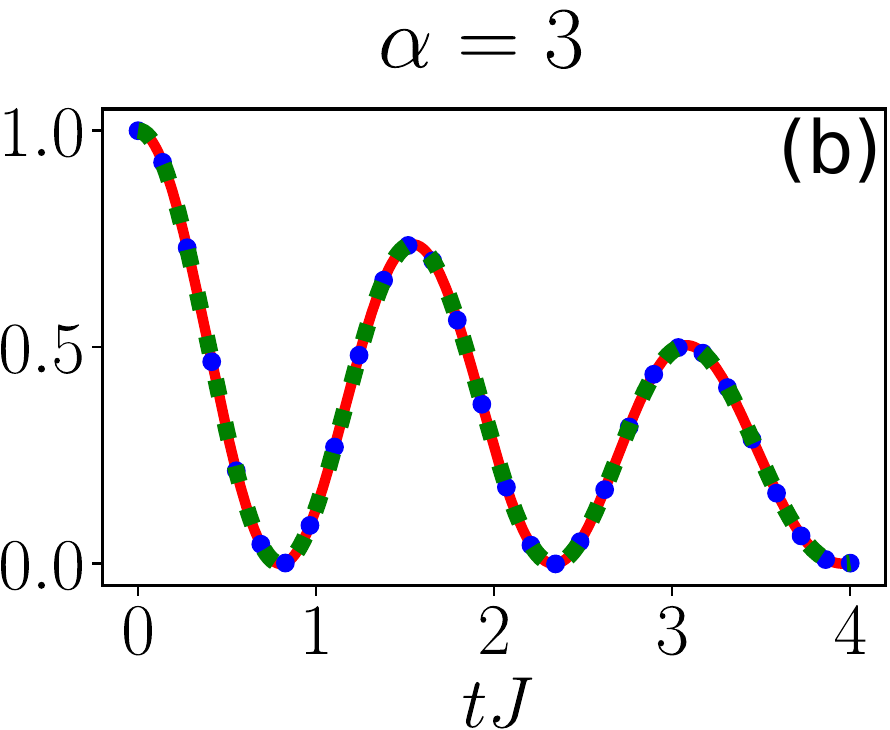}  & \includegraphics[width = 55mm]{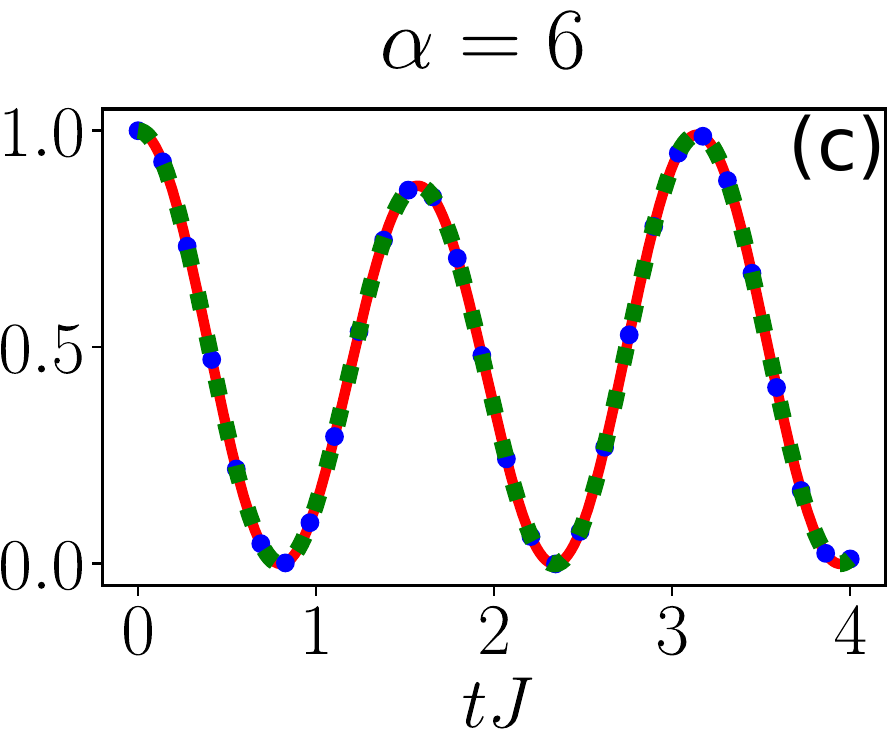}      \\ 
  \includegraphics[width = 58.8mm]{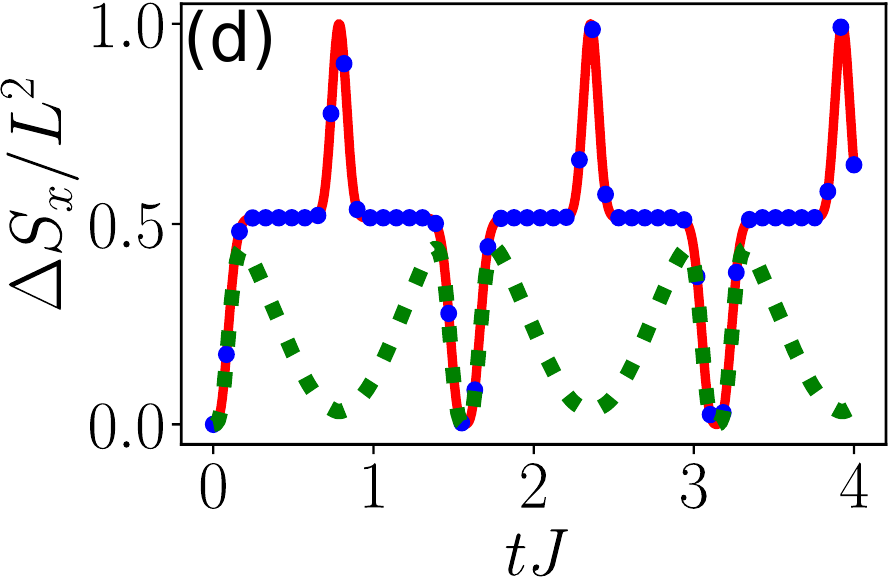} & \includegraphics[width = 55mm]{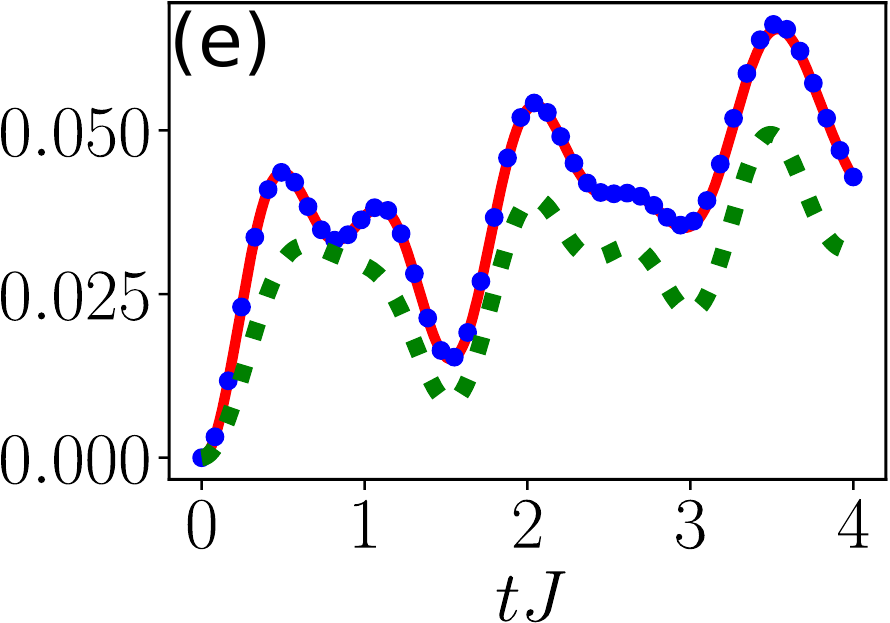}  & \includegraphics[width = 55mm]{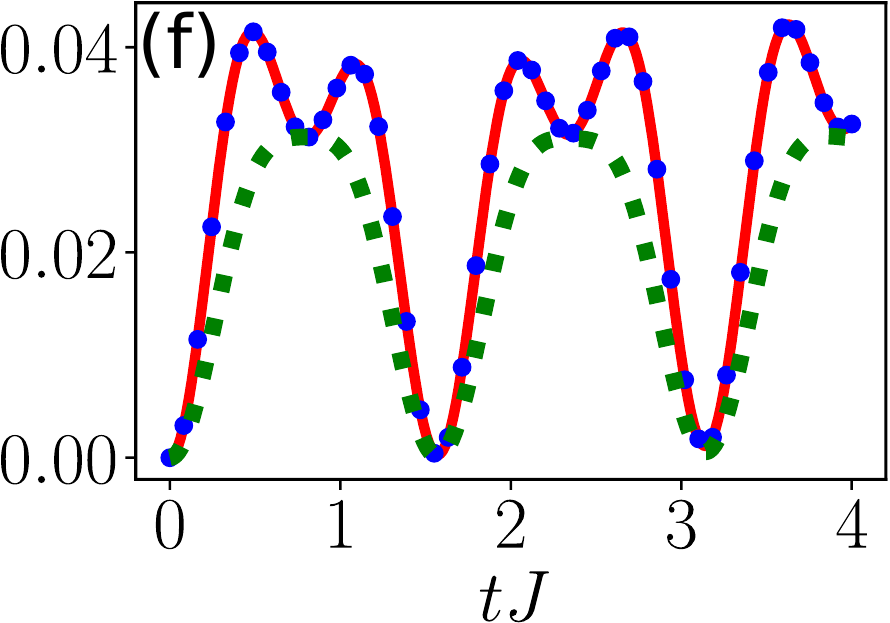}  
  \\ 
    \includegraphics[width = 56.8mm]{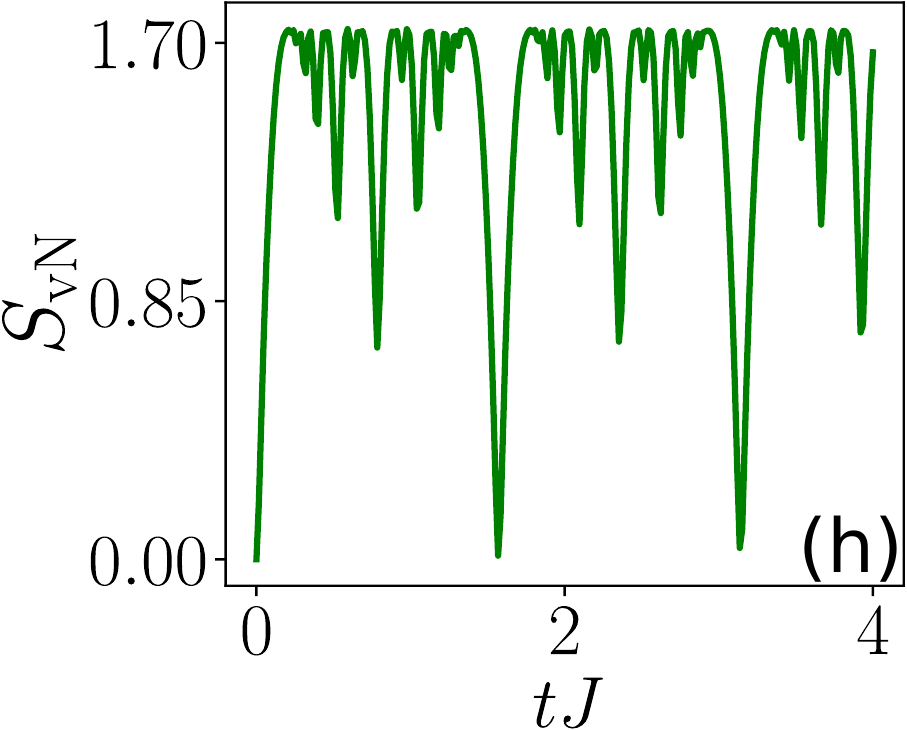} & \includegraphics[width = 55mm]{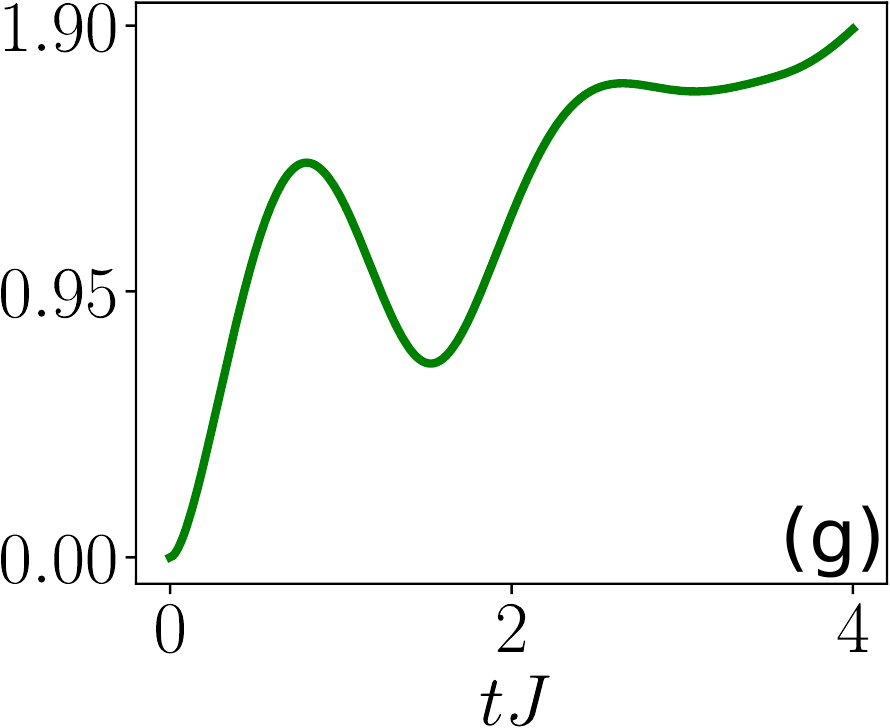}  & \includegraphics[width = 55mm]{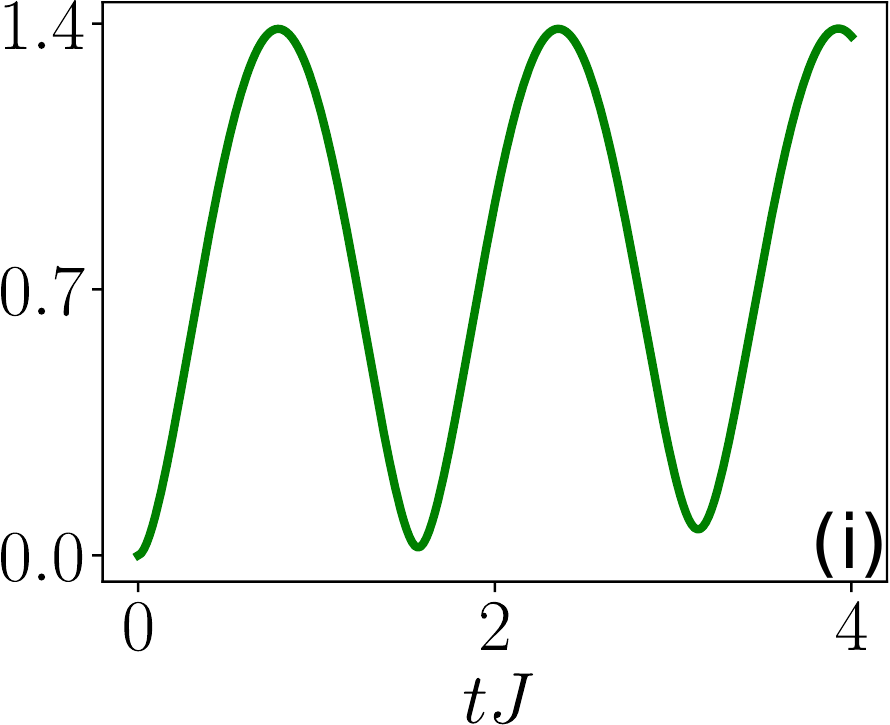} 
\end{tabular}
\end{center}
\caption{The dynamics of the Ising model in a 1D lattice of size $L=32$ is shown. Time evolution of the expectation value of the collective spin operator $\hat{S}_x $ in (a--c) and the collective correlation function $\Delta \hat{S}_x$ in (d--f) are depicted for the interaction ranges with all-to-all ($\alpha=0$), dipolar ($\alpha=3$), and Van der Waals ($\alpha=6$) scaling, respectively. (h--i) illustrate the time evolution of the entanglement entropy for a quarter-chain bipartition, computed using ML-MCTDH, based on the tree structure outlined in Appendix \ref{MLMCTDH}.} 
\label{IsingResult}
\end{figure*}
\subsection{Spin models}
\label{spin_models}
 We study the dynamics of the Heisenberg model in the Ising and anisotropic XYZ limits to benchmark the performance of the ML-MCTDH. We also consider long-range couplings, which are particularly important as many quantum simulators naturally give rise to such interactions \cite{Yan2013,PhysRevLett.113.195302,PhysRevX.11.011011,Scholl2021,Ebadi2021,Britton2012,RevModPhys.93.025001}. We examine numerically challenging edge cases such as the disordered Ising model with random couplings as well as all-to-all interactions to explore the limitations of the method. 
 \begin{figure*}
\begin{center}
\begin{tabular}{ccc}
  \includegraphics[width=55mm]{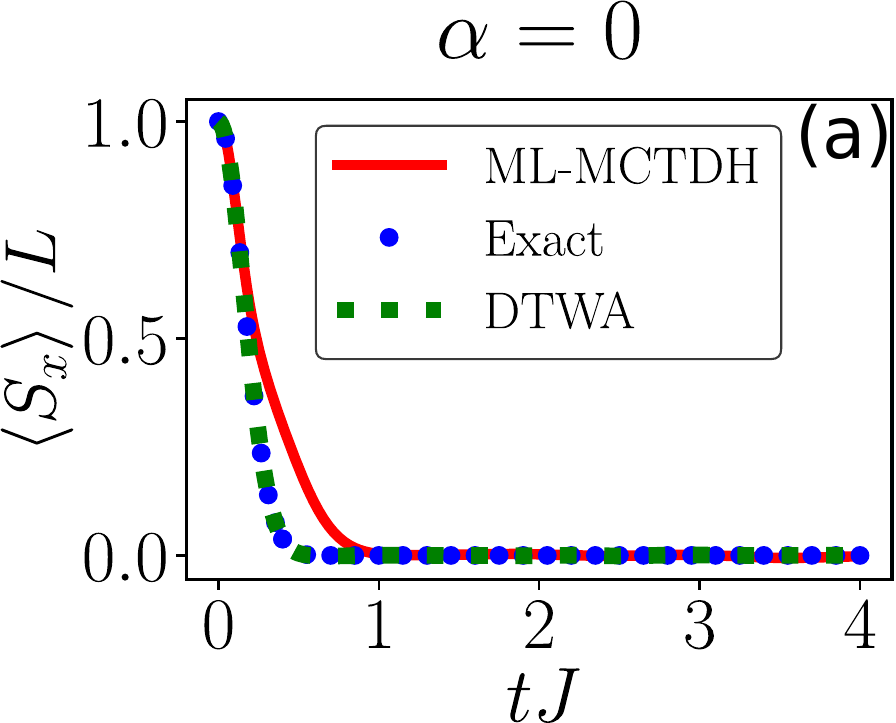}   & \includegraphics[width=55mm]{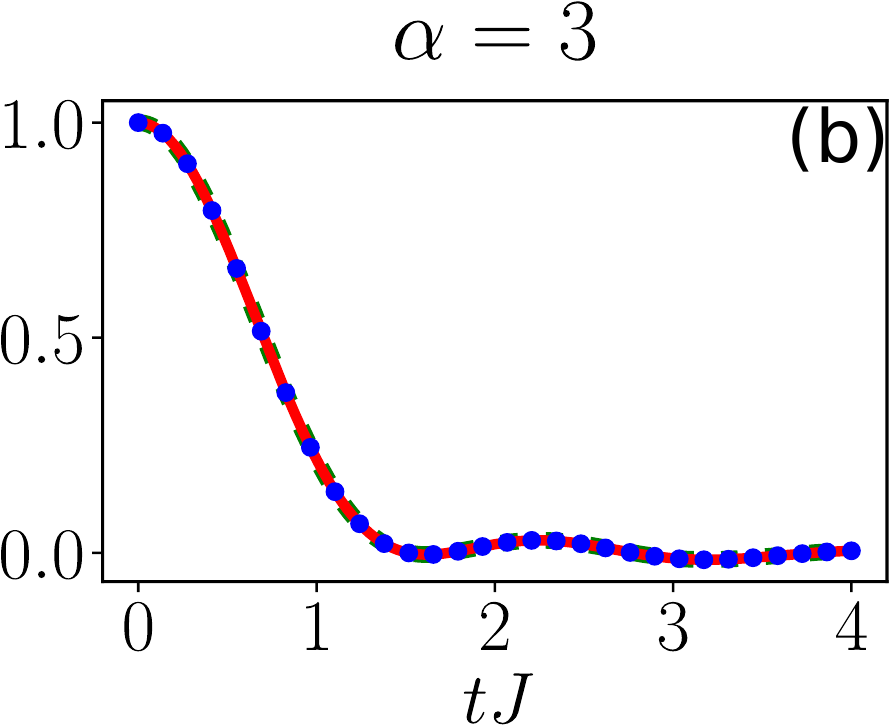} & \includegraphics[width = 55mm]{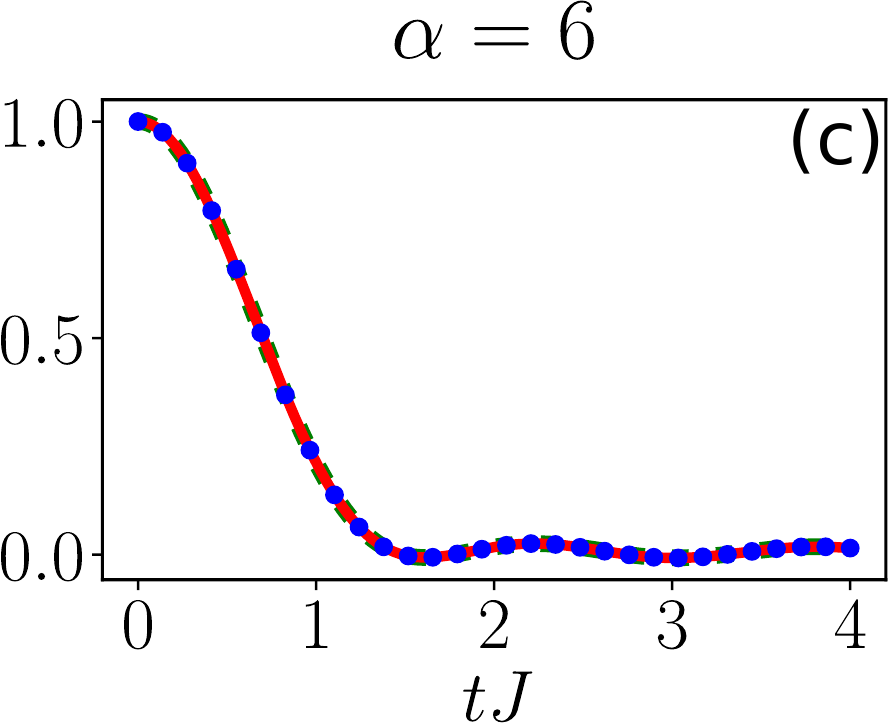}      \\
  \includegraphics[width = 55mm]{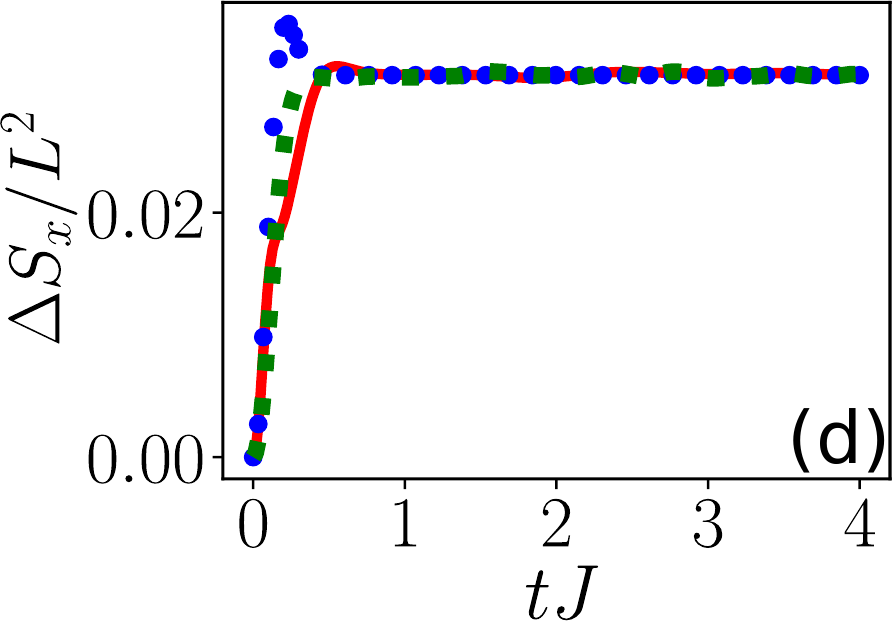} & \includegraphics[width = 55mm]{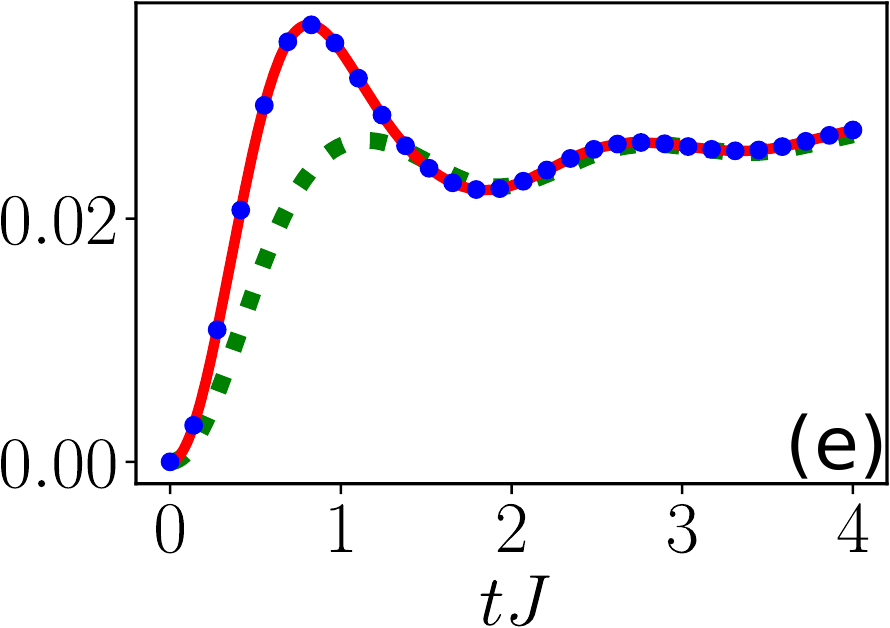} & \includegraphics[width = 55mm]{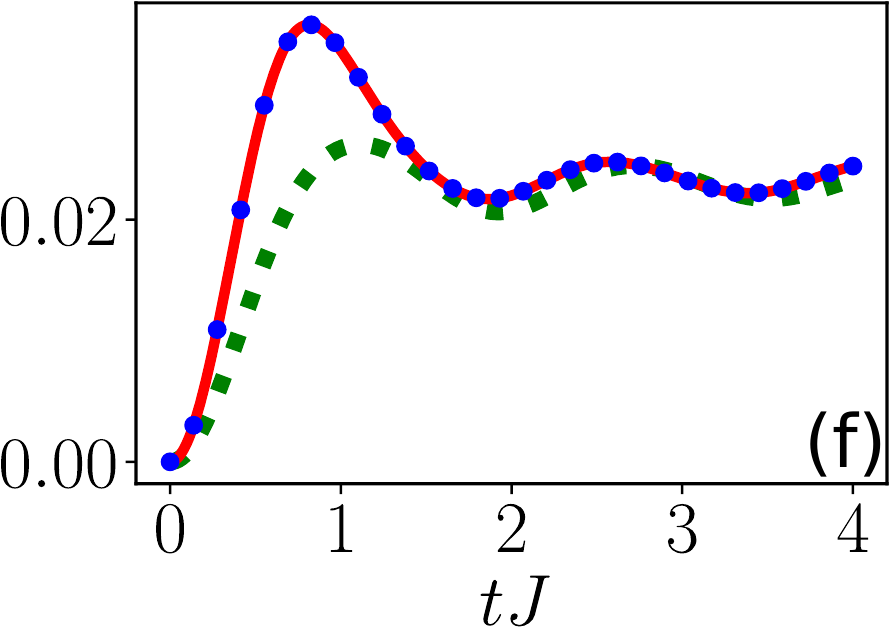}   
\end{tabular}
\end{center}
\caption{Dynamics of the disordered Ising model ($J_{x,y}=0$), with $J_z = J_{ij}/J$ in 1D lattice of size, $L=32$ is shown. Here, $J_{ij}$ represents the interaction strength between spins located at sites $i$ and $j$, and is chosen randomly from a uniform distribution within the range $[-1, 1]$. We averaged over {100} disorder realizations.  (a--c) Time evolution of expectation value of collective spin $ \hat{S}_x $, for all-to-all ($\alpha=0$), dipolar ($\alpha=3$) and van der Waals ($\alpha=6$) interactions. (d--f) Time evolution of the collective correlation function $\Delta \hat{S}_x$, for all-to-all ($\alpha=0$), dipolar ($\alpha=3$) and van der Waals ($\alpha=6$) interactions. } 
\label{DisorderedResult}
\end{figure*}
 
 The Heisenberg Hamiltonian for a system with $L$ spins is given by
\begin{equation} 
    \hat{H} =  -\sum_{\substack{i,j=1\\i<j}}^{L} \left[ J_{ij}^{x} \hat{\sigma}_{i}^x \hat{\sigma}_{j}^x + J_{ij}^{y} \hat{\sigma}_{i}^y \hat{\sigma}_{j}^y + J_{ij}^{z} \hat{\sigma}_{i}^z \hat{\sigma}_{j}^z \right],
    \label{XYZHamiltonian}
\end{equation}
where $J_{i,j}^{\beta}=J_{\beta}/r_{ij}^{\alpha}$ represents the interaction strength along the $\beta=x,y,z$ axis between spins residing at sites $i$ and $j$. The interactions decay as a function of the distance $r_{ij}$ with an exponent $\alpha$, which controls the range of the interaction. Decay exponents relevant to various quantum simulation platforms are considered, including polar molecules ($\alpha =3$) \cite{Yan2013,PhysRevLett.113.195302}, Rydberg atoms ($\alpha =3,6$) \cite{PhysRevX.11.011011,PhysRevLett.120.063601,Scholl2021,Ebadi2021,Ryd_toric,Qsimfrust}, and trapped ions ($\alpha = 0-3$) \cite{Britton2012,RevModPhys.93.025001}. Nearest-neighbor interactions are also examined.

Exact analytical expressions for the dynamics exist in the Ising limit $(J^x,J^y=0)$ \cite{Worm_2013,PhysRev.107.46}, rendering it an ideal benchmark for the time evolution of large spin systems that go beyond the limits of exact diagonalization. Therefore, we start with the Ising model as it is also one of the most fundamental and well-studied models \cite{RevModPhys.39.883} and has been realized in experiments with ion traps \cite{Jurcevic2014,Britton2012,Richerme2014,PhysRevLett.92.207901,PhysRevLett.103.120502,doi:10.1126/science.1232296} and Rydberg atoms \cite{Weimer2010,doi:10.1126/science.1258351,PhysRevLett.106.025301,Labuhn2016,PhysRevResearch.3.013286}. 

Following this, the XYZ model ($J^x \ne J^y \ne J^z$) is considered. 
Compared to the Ising model, the XYZ model is more complex due to the presence of non-commuting terms and anisotropy in the Hamiltonian, which also has been recently realized in experiments \cite{luo2024hamiltonianengineeringcollectivexyz,doi:10.1126/science.abd9547,Miller2024,PhysRevApplied.9.064029}. 

Experimental realizations of many-body quantum systems often exhibit inherent disorder \cite{doi:10.1143/JPSJ.35.1593,OSHIKAWA1997533,HolgerFrahm_1997,PhysRevLett.48.1559,PhysRevLett.50.1395,RevModPhys.71.875} and are known to violate the area law of entanglement entropy  \cite{RevModPhys.82.277,PhysRevB.97.125116,PhysRevA.99.052342,PhysRevA.71.012301,PhysRevA.93.053620}, which can pose challenges for MPS-based methods \cite{Orús2019,RevModPhys.82.277,Vitagliano_2010,PhysRevLett.93.260602,PhysRevB.102.014455}. A recent study \cite{PhysRevResearch.5.023135} has shown that ML-MCTDH is capable of accurately describing ground state properties of disordered spin systems. Here, we simulate the dynamics of a disordered system, specifically the
disordered Ising model ($J^x,J^y=0$) with $J^z = J_{ij}$ chosen randomly from $\left[-1,1\right]$.

\section{Results}
\label{Results}
We demonstrate the capability of ML-MCTDH to accurately depict the time evolution of various spin models characterized by different interaction ranges. This is illustrated by calculating the dynamics of the expectation value of the collective spin $\braket{\hat{S}_x}=\sum_{i} \braket{\hat{\sigma}_i^x}$ and correlation function $\Delta \hat{S}_x= \braket{\hat{S}_x^2} - \braket{\hat{S}_x}^2 = \sum_{i,j}\braket{\hat{\sigma}_i^x \hat{\sigma}_j^x} -\braket{\hat{\sigma}_i^x}\braket{\hat{\sigma}_j^x}$. For the Ising model, exact analytical expressions are available for the dynamics \cite{Worm_2013,PhysRev.107.46} and are used to benchmark the performance of ML-MCTDH. For the XYZ model, exact diagonalization (ED) is employed due to the absence of such analytical results. Across all spin models, a comparative analysis evaluates the performance of ML-MCTDH alongside DTWA. The initial state for all simulations is the paramagnetic state, denoted as 
$\ket{\rightarrow\rightarrow\cdots\rightarrow}$, which represents equal superposition of all the $\sigma_z$ basis states. { This corresponds to the ground state of the transverse field Ising model in the limit of an infinite transverse field. We then perform a sudden quench to different limits, all with zero transverse field, and analyze the post-quench behavior of one-body and two-body correlators. We investigate the behavior of the entanglement dynamics to understand the underlying post-quench dynamics and obtain information about the complexity of the quantum states. The entanglement is quantified by computing the bipartite von Neumann entropy $S_{\text{vN}}\equiv -\Tr(\rho_r\ln{\rho_r})$ of the states over time, where $\rho_r$ is the reduced density matrix of a bipartition of the system.} 
 \begin{figure*}
\begin{center}
\begin{tabular}{ccc}
  \includegraphics[width=55mm]{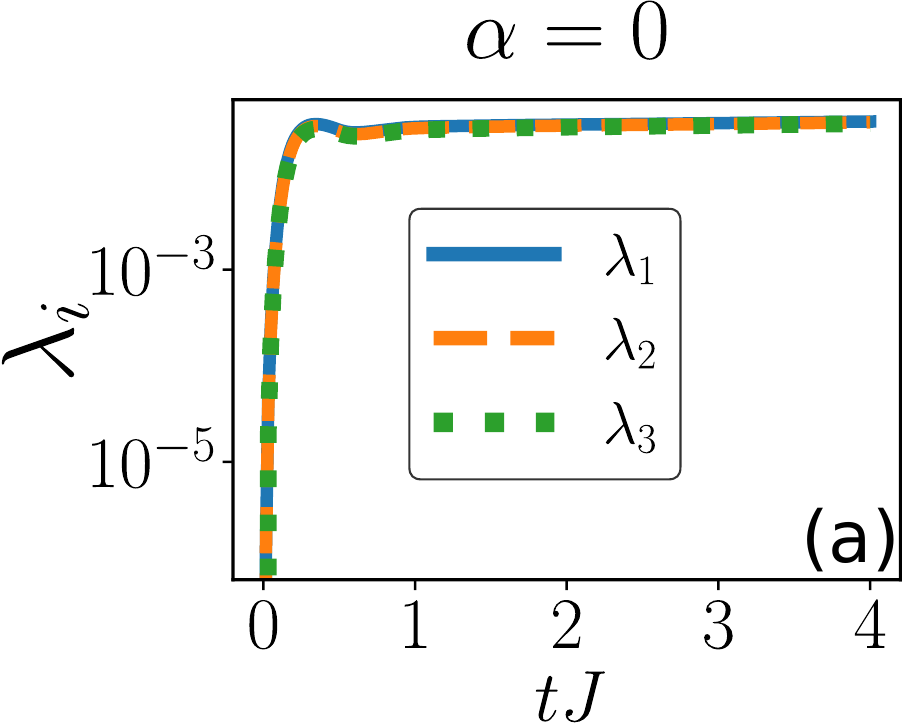}   & \includegraphics[width=55mm]{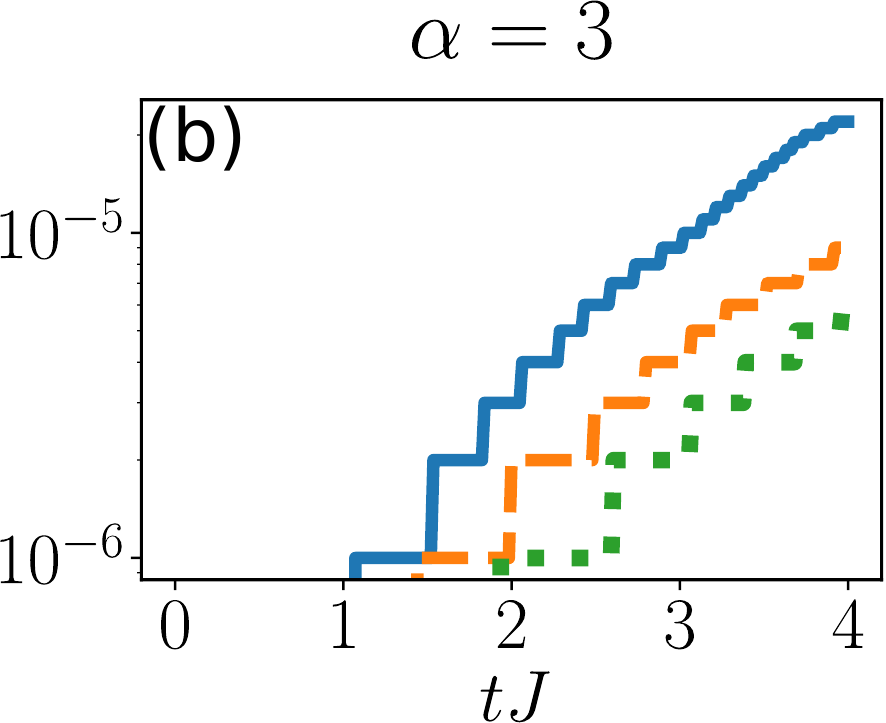} & \includegraphics[width = 55mm]{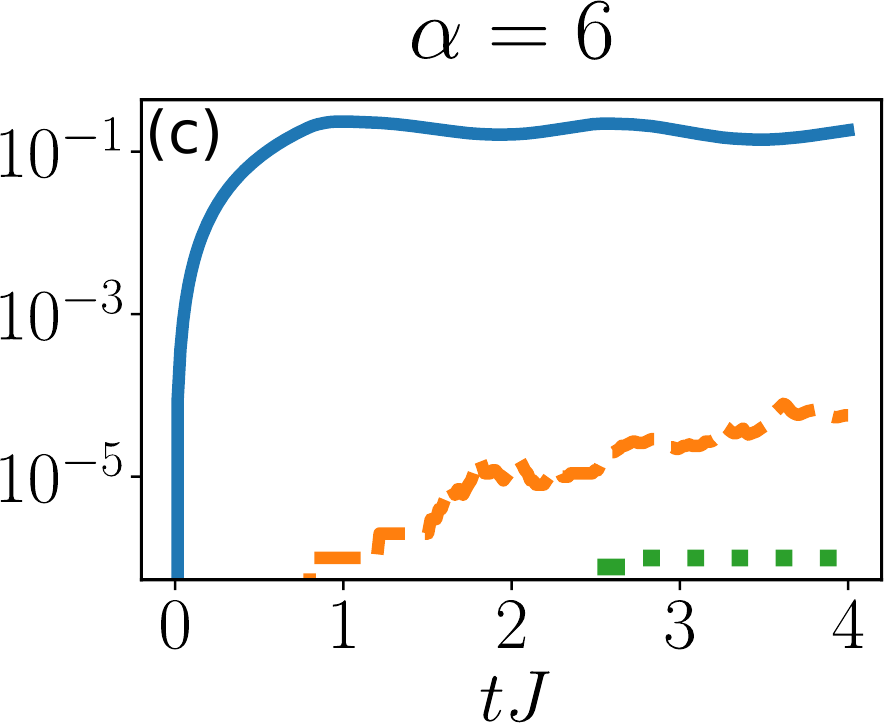}      \\
  \includegraphics[width = 55mm]{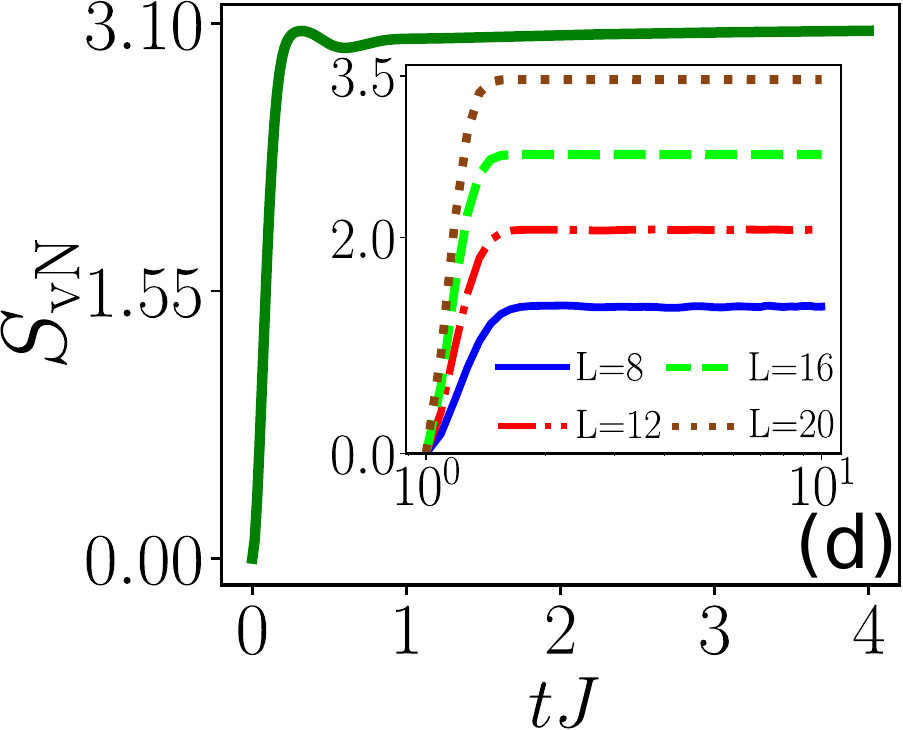} & \includegraphics[width = 55mm]{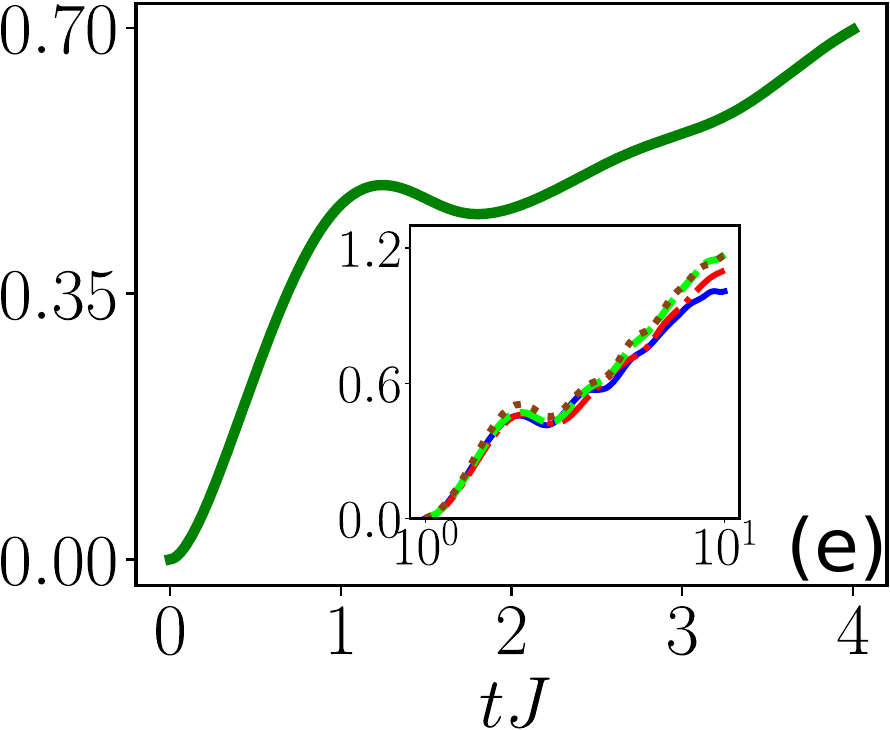} & \includegraphics[width = 55mm]{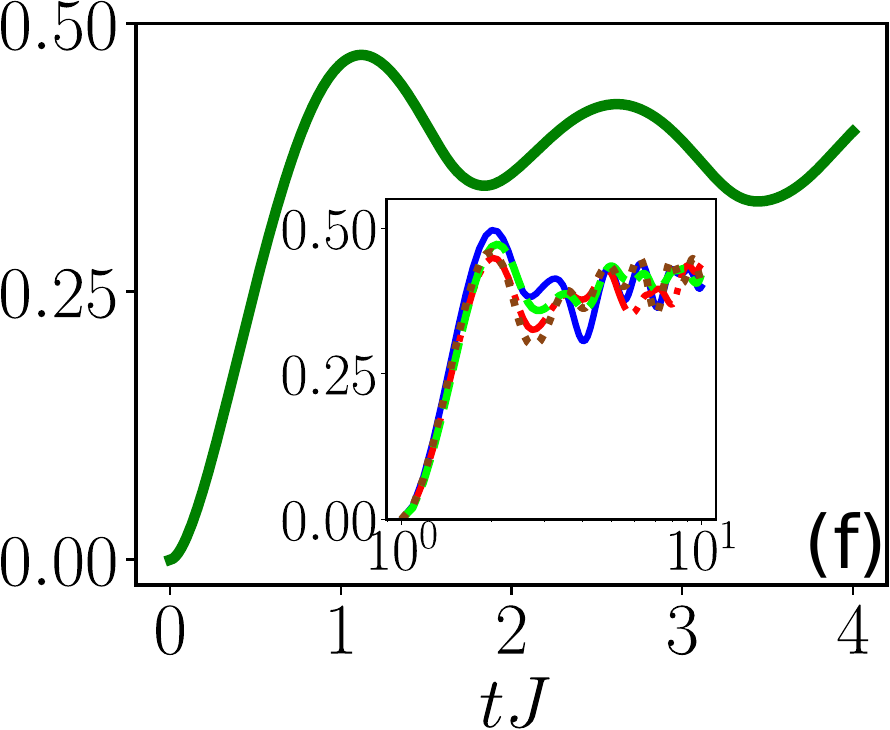}   
\end{tabular}
\end{center}
\caption{Dynamics of the disordered Ising model ($J_{x,y}=0$), with $J_z = J_{ij}/J$ in 1D lattice of size, $L=32$ is shown. Here, $J_{ij}$ represents the interaction strength between spins located at sites $i$ and $j$, and is chosen randomly from a uniform distribution within the range $[-1, 1]$. We averaged over 100 disorder realizations.  (a--c) The three least dominant natural population during the evolution, for all-to-all ($\alpha=0$), dipolar ($\alpha=3$) and van der Waals ($\alpha=6$) interactions. (d--f) The time evolution of the entanglement entropy for a quarter-chain bipartition, computed using ML-MCTDH, based on the tree structure outlined in Appendix \ref{MLMCTDH}. Insets show the ED results for time evolution of entanglement entropy for a quarter-chain bipartition for different sizes and interaction ranges. } 
\label{DisorderedEnt}
\end{figure*}



In the Ising limit $(J^x,J^y=0)$, the time evolution of the operators $\hat{S}_x$ and $\Delta \hat{S}_x$ for long-range interactions is shown in Fig.~\ref{IsingResult}. In the all-to-all interacting case ($\alpha=0$), the dynamics of the observable $\braket{\hat{S}_x}$ exhibits revivals at times determined by the multiples of $\pi/2J$ as can be deduced from the exact analytical results. ML-MCTDH captures this behavior exactly throughout the entire simulation as shown in Fig.~\ref{IsingResult}(a). In the dipolar ($\alpha=3$) and Van der Waals ($\alpha=6$) interacting cases, $\braket{\hat{S}_x}$ displays oscillatory dynamics and is exactly reproduced by the ML-MCTDH as can be seen in Fig.~\ref{IsingResult}(b) and (c), respectively. Equally, DTWA reproduces the corresponding dynamics of $\braket{\hat{S}_x}$ correctly. For the correlation ${\Delta \hat{S}_x}$, however, DTWA only manages to reproduce the dynamics for a short time around each time point that are multiples of $\pi/2J$ for $\alpha=0$ while ML-MCTDH reproduces the revival dynamics exactly at all times as shown in Fig.~\ref{IsingResult}(d). In the case of $\alpha = 3,6$, the time evolution of ${\Delta \hat{S}_x}$ is fully reproduced by ML-MCTDH while DTWA captures the trend only qualitatively as can be seen in Fig.~\ref{IsingResult}(e) and (f). Even the very minute variations in the behavior of ${\Delta \hat{S}_x}$ are accounted for with ML-MCTDH while DTWA tends to smoothen out such small changes occurring over short time intervals as can be seen in Fig.~\ref{IsingResult}(e) and (f) at $tJ \in [0,1.5]$. { In long-range interacting models, non-local interactions allow quantum correlations to form rapidly between distant degrees of freedom. However, despite this rapid correlation growth, the entanglement dynamics, as depicted in Fig.~\ref{IsingResult} (h--i), following a quench exhibits an oscillatory behavior for interactions with $\alpha= 0,6$ while the entanglement entropy increases linearly over time for interactions with $\alpha = 3$.}

\begin{figure*}
\begin{center}
\begin{tabular}{ccc}
  \includegraphics[width=55mm]{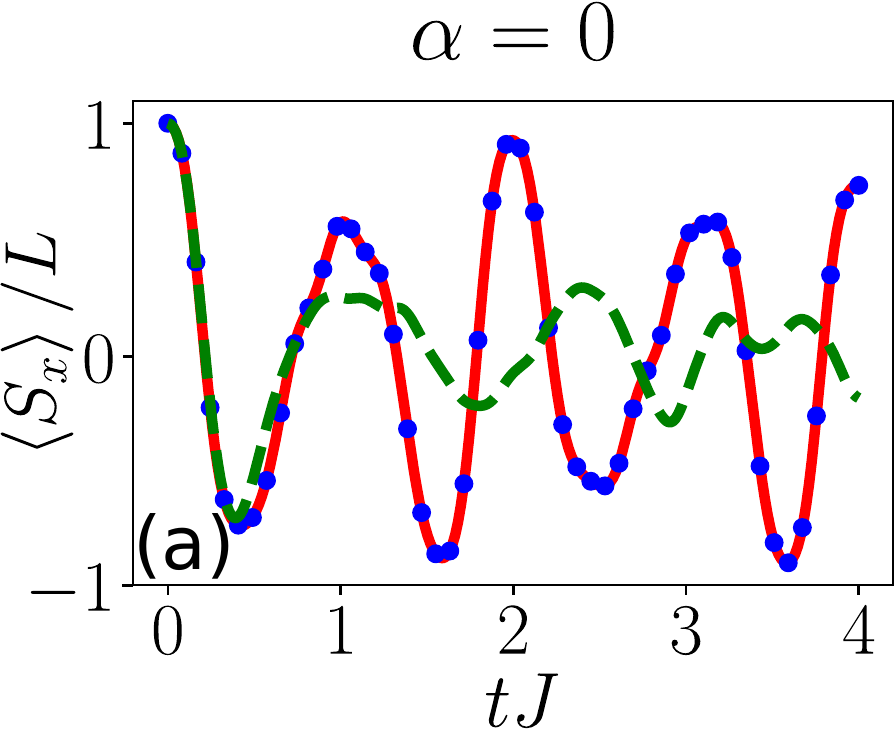} & \includegraphics[width=55mm]{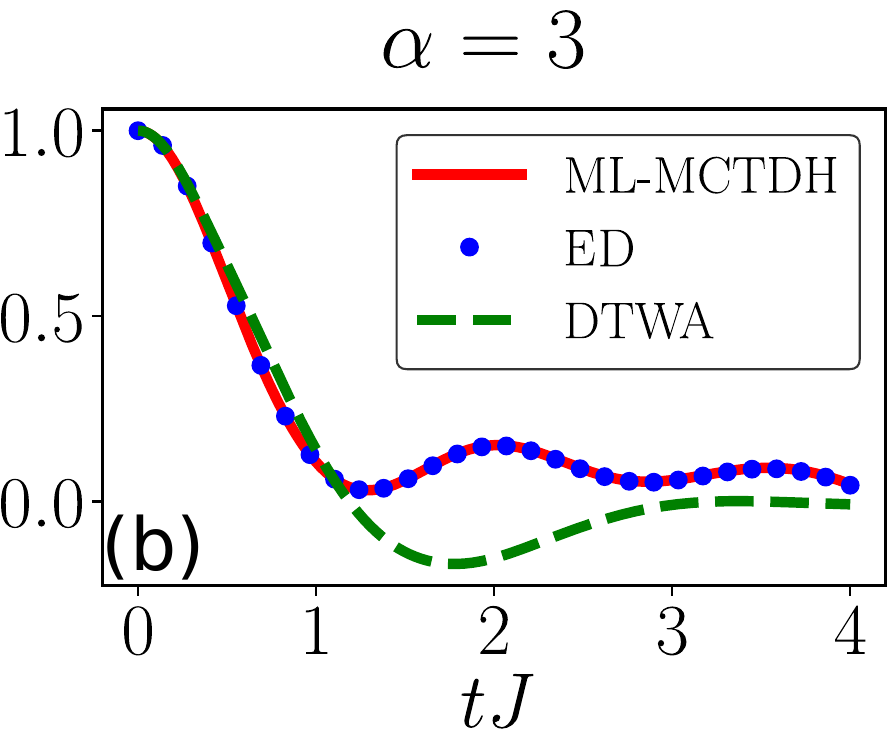} & \includegraphics[width = 55mm]{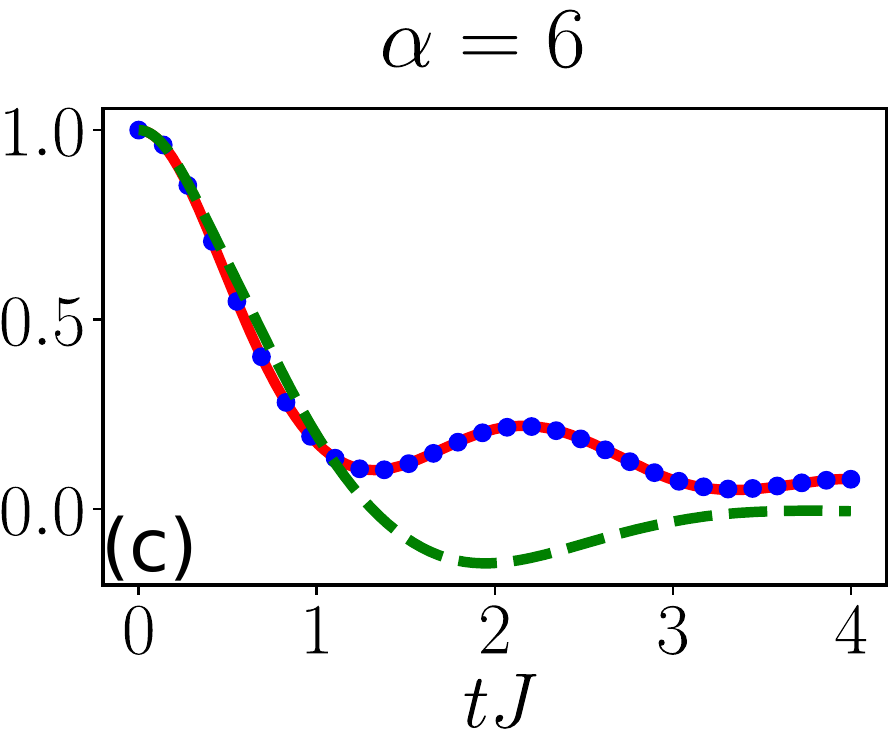}     \\
  \includegraphics[width = 55mm]{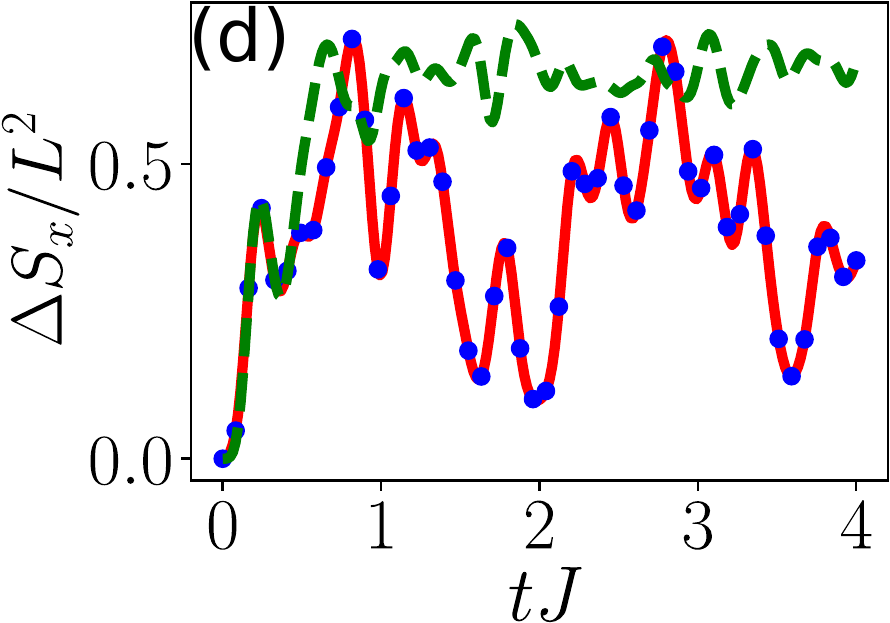} & \includegraphics[width = 55mm]{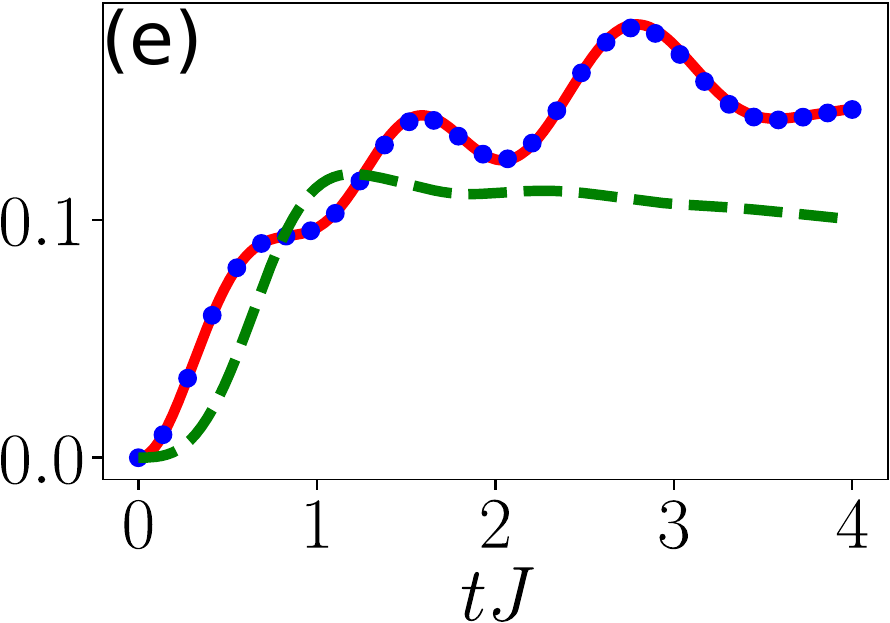}  & \includegraphics[width = 55mm]{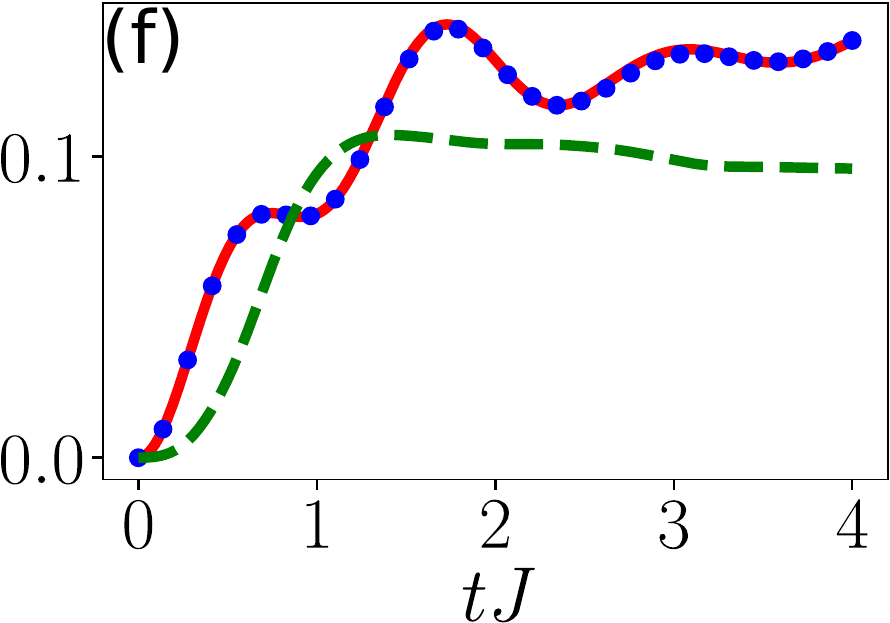}    
\end{tabular}
\end{center}
\caption{Dynamics of the XYZ model with coupling strengths $J_x = 0.5, J_y(\equiv J) = 1.0, J_z = 0.25$ in a 1D lattice of size $L=16$ is shown. Time evolution of the expectation value of the collective spin operator $\hat{S}_x $ in (a--c) and the collective correlation function $\Delta \hat{S}_x$ in (d--f) are displayed for the interaction ranges with all-to-all ($\alpha=0$), dipolar ($\alpha=3$), and Van der Waals ($\alpha=6$) scaling, respectively. } 
\label{XYZResult}
\end{figure*}


For the disordered Ising model, the time evolution of the operators $\hat{S}_x$ and $\Delta \hat{S}_x$ is shown in Fig.~\ref{DisorderedResult}. The choice of a uniform distribution in $[-1, 1]$ for $J_{ij}$ allows the presence of both ferromagnetic ($J_{ij} > 0$) and antiferromagnetic ($J_{ij} < 0$) interactions in the system which makes the calculation of dynamics challenging. In the $\alpha=0$ case, $\braket{\hat{S}_x}$ undergoes a rapid decay. ML-MCTDH captures this initial decay with minor deviations and converges to the exact results for longer times. Fig.~\ref{DisorderedResult}(b) and (c) show that ML-MCTDH reproduces the exact analytical solution of $\braket{\hat{S}_x}$  for $\alpha=3$ and $\alpha=6$. $\Delta\hat{S}_x$ depicts a rapid initial growth and quickly reaches saturation in the $\alpha=0$ case, as shown in Fig.~\ref{DisorderedResult}(d). Both ML-MCTDH and DTWA closely approximate the initial rise and align with the exact results after a short time. In cases of $\alpha=3$ and $\alpha=6$, ML-MCTDH replicates the exact analytical solution for the dynamics of $\Delta\hat{S}_x$, as depicted in Fig.~\ref{DisorderedResult}(e) and (f). However, DTWA misses the initial rise, although it closely matches the exact results at later stages. 

The ML-MCTDH approach is a numerically exact method, meaning that the convergence of results can be systematically verified by increasing the number of time-dependent basis states. Therefore, the deviations observed in ML-MCTDH simulations of disordered Ising for $\alpha=0$ in Fig.~\ref{DisorderedResult} (a,d) can be attributed to having insufficient time-dependent basis states. This can be confirmed by analyzing the natural populations of the wavefunction. The natural populations correspond to the eigenvalues of the one-body density matrices and provide a measure of the effectiveness of the chosen basis set. If only a few natural populations dominate, this indicates that a reduced number of basis states describe the system well. This is evident in Fig.~\ref{DisorderedEnt}(b,c) for $\alpha=3,6$, where the least dominant natural population stays below $10^{-5}$ throughout the evolution. Furthermore, the populations decrease sharply compared to the next larger one, highlighting a non-uniform distribution. Conversely, a more uniform distribution suggests the need for a larger basis set to capture the system's complexity accurately. This is reflected in Fig.~\ref{DisorderedEnt} for $\alpha=0$, where the least dominant values remain relatively large and do not exhibit any drop-off, indicating that a larger set of basis states is required to represent the wavefunction accurately. The deviations observed can be explained through analyzing the behavior of the entanglement dynamics. In the long-range Ising disordered model the entanglement dynamics exhibit a stark contrast to the case without disorder. For $\alpha=0$, entanglement entropy increases rapidly before saturating in a size-dependent manner, approaching its theoretical maximum of $\text{ln}(2^{L/4})$, as shown in the inset of Fig. \ref{DisorderedEnt}(d). While ML-MCTDH successfully captures this rapid entanglement growth (Fig. \ref{DisorderedEnt}(d)), its limited number of time-dependent states prevents it from fully reproducing the correct entanglement. Therefore, the number of time-dependent basis states must be set close to the maximum to accurately capture the wavefunction. This rapid entanglement growth significantly challenges classical simulability, even for short-time dynamics, necessitating a dedicated analysis of SPF scaling and the tree structure for optimal state representation. For $\alpha=3$, the entanglement entropy 
shows a logarithmic growth after a short rapid initial growth, as shown in the inset of Fig. \ref{DisorderedEnt}(e). ML-MCTDH captures this behavior well with a relatively small number of time-dependent basis states \ref{DisorderedEnt}(e). In the $\alpha=6$ case, the entanglement entropy
becomes independent of system size, exhibiting only a brief early growth. This suggests that a limited number of basis states are sufficient to accurately capture the dynamics over arbitrarily long times \cite{PhysRevB.95.094205}.


\begin{figure*}
 \centering
\begin{tabular}{ccc}
  \includegraphics[width=56mm]{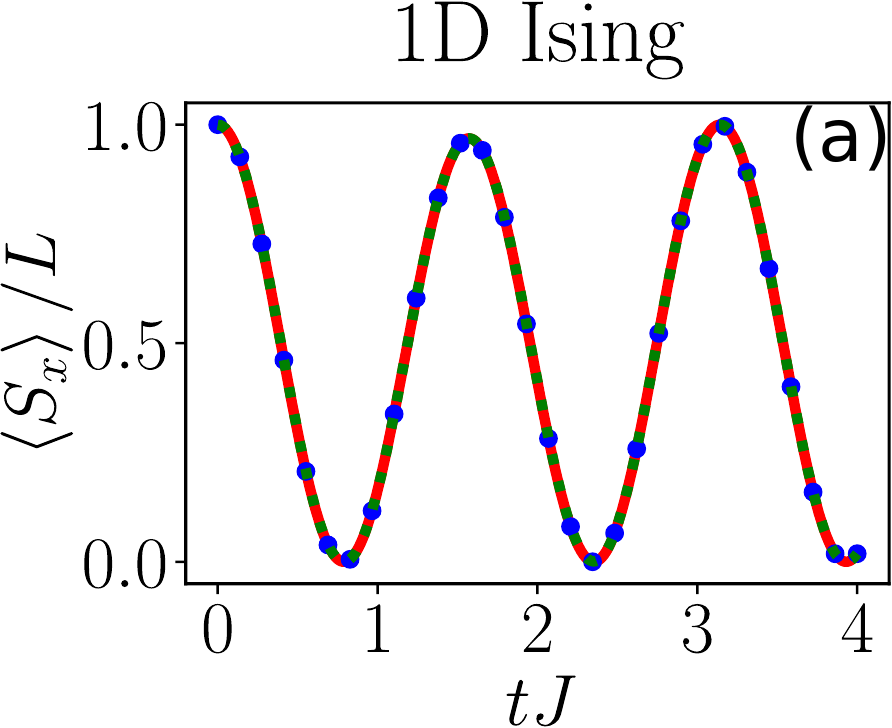}& \includegraphics[width = 55mm]{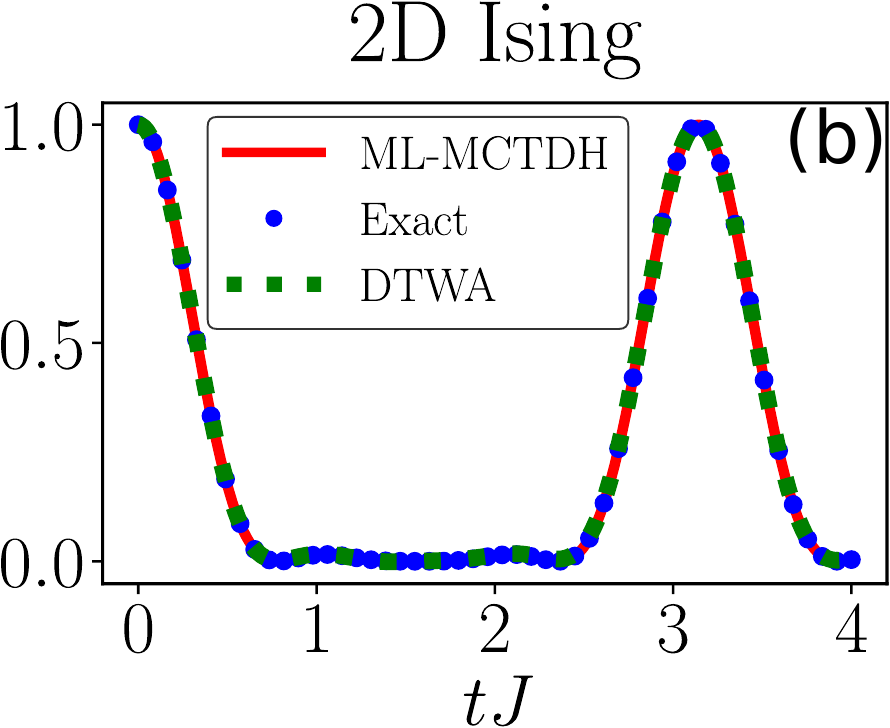}  & \includegraphics[width = 55mm]{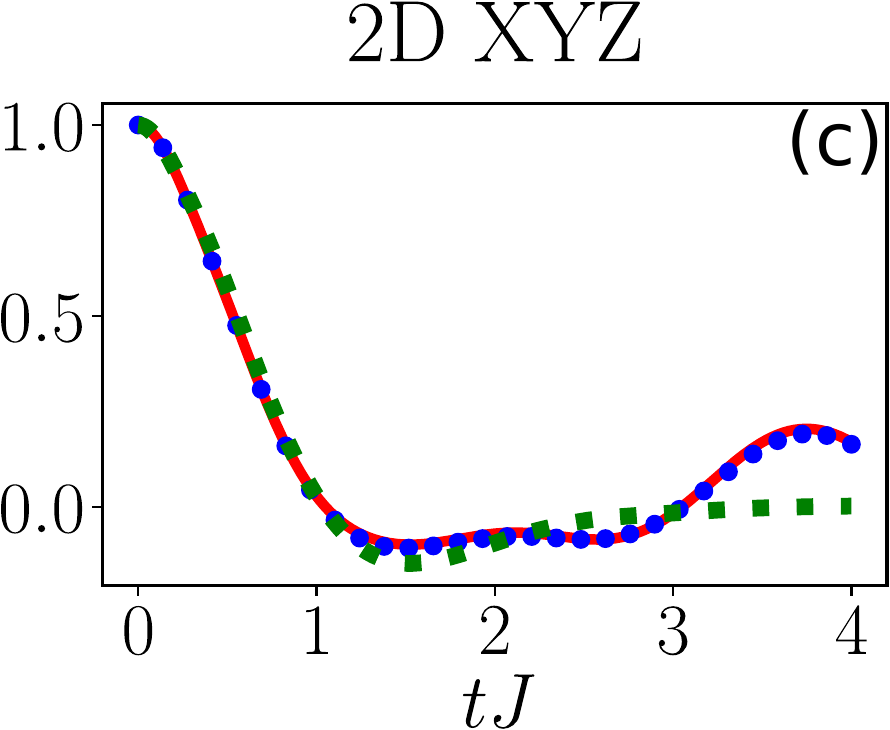}      \\
  \includegraphics[width = 56mm]{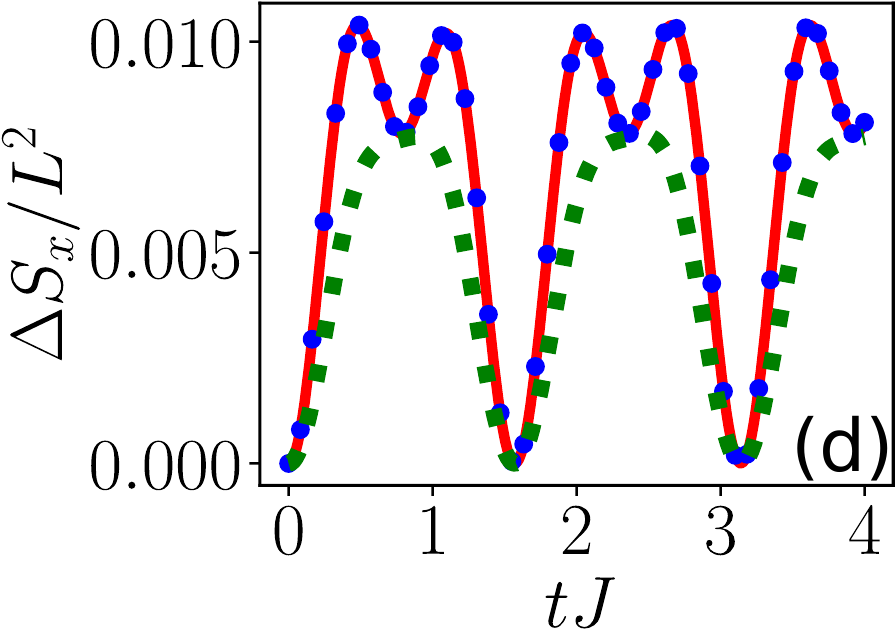} & \includegraphics[width = 55mm]{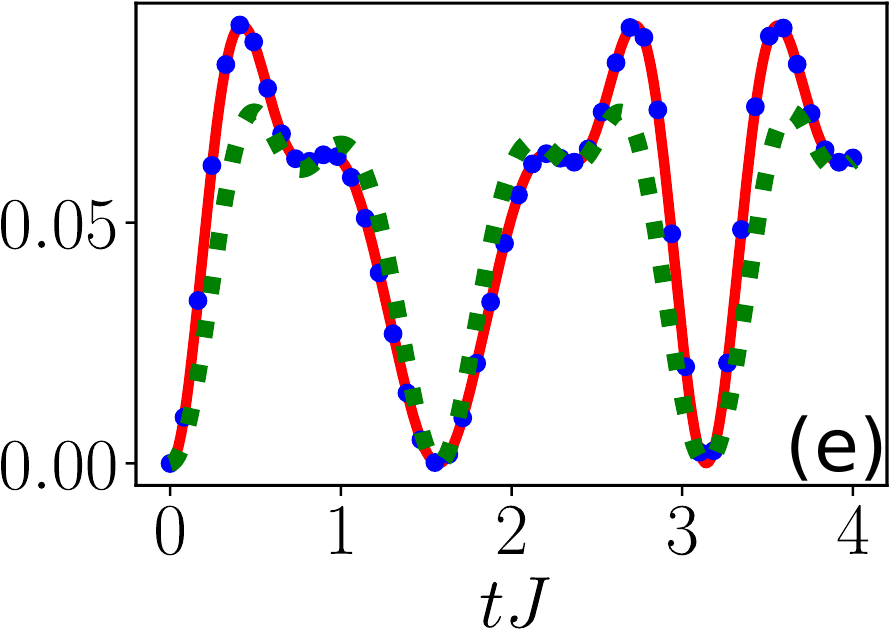}  & \includegraphics[width = 55mm]{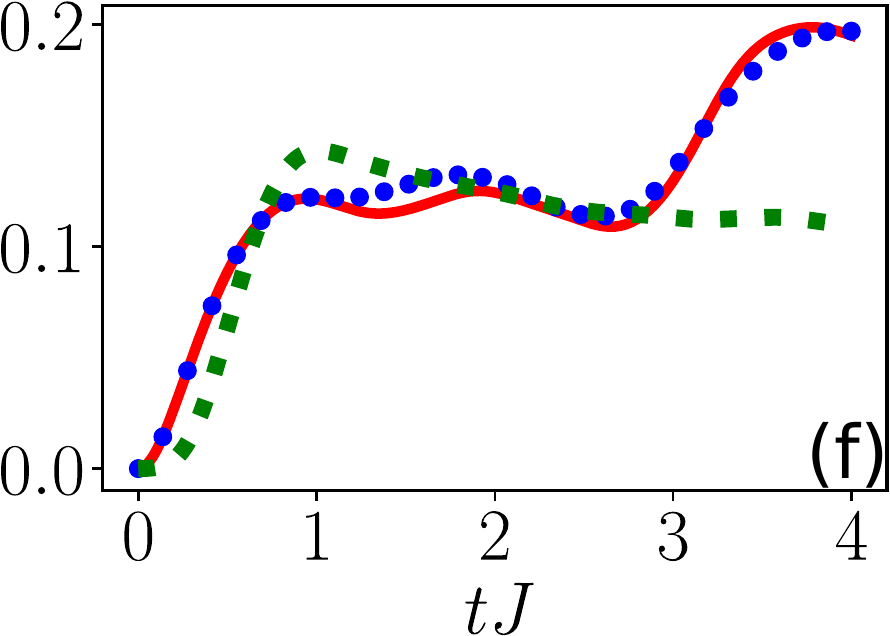}      
\end{tabular}
\caption{The dynamics of the 1D Ising model ($L=128$), 2D Ising model ($L = 4 \times 4$), and 2D XYZ model ($L= 4 \times 4$) with coupling strengths $J_x = 0.5$, $J_y (\equiv J) = 1.0$, and $J_z = 0.25$, considering nearest-neighbor interactions, are shown. In all plots, the red solid line represents the ML-MCTDH results, the blue circles denote the exact results, and the green squares correspond to the DTWA results. The exact results refer to analytical solutions for the Ising interaction and exact diagonalization for the XYZ interaction. Panels (a--c) display the time evolution of the expectation value of the collective spin operator $\hat{S}_x$, while panels (d--f) show the dynamics of collective correlation function $\Delta \hat{S}_x$ for each model.} 
\label{NNResult}
\end{figure*}

Fig.~\ref{XYZResult} displays the dynamics of observables $\braket{\hat{S}_x}$ and $\Delta \hat{S}_x$ in the XYZ case. For $\alpha=0$, ML-MCTDH reproduces the exact dynamics of $\braket{\hat{S}_x}$ as obtained from ED calculations. DTWA can only capture the short-time behavior as shown in Fig.~\ref{XYZResult}(a). For $\alpha=3,6$, ML-MCTDH matches the ED results for $\braket{\hat{S}_x}$ throughout the entire dynamics while DTWA only captures the initial decay, as shown in Fig.~\ref{XYZResult}(b) and (c). For $\alpha=0$, ML-MCTDH accurately captures the dynamics of ${\Delta \hat{S}_x}$ throughout the entire simulation as shown in Fig.~\ref{XYZResult}(d), particularly correctly describing the sudden changes in the behavior over short time intervals. In contrast, DTWA is accurate only at the very early stages of the dynamics and quickly saturates to incorrect values. For $\alpha = 3,6$, ML-MCTDH results overlap with exact dynamics of ${\Delta \hat{S}_x}$, accurately capturing the increasing trend in correlations as can be seen in Fig.~\ref{XYZResult}(e) and (f). DTWA only matches the qualitative behavior at short times, incorrectly indicating a decrease in correlations at later stages in the dynamics. The overall tendency of DTWA to be accurate at only short times for the XYZ model can be attributed to the fact that it only accounts for the leading-order quantum corrections to a mean-field approximation \cite{PhysRevB.93.174302,PhysRevResearch.3.013060}. Therefore, the validity of the timescale for the DTWA heavily depends on the model and the range of interactions. On the other hand, independent of the model, ML-MCTDH specifically attempts to account for the beyond mean-field correlations by employing an optimal number of time-dependent basis states. {In the XYZ limit, the growth of entanglement exhibits a behavior similar to that of the Ising model. A wavefunction with higher entanglement entropy spans a large number of basis state configurations, requiring a greater number of SPFs for accurate representation (see Appendix \ref{MLMCTDH}).}

We evaluate the performance of ML-MCTDH on the scalability and efficiency in dealing with high dimensional lattices by studying the dynamics of the long 1D Ising model alongside 2D Ising and XYZ models as depicted in Fig.~\ref{NNResult}. DTWA matches the exact dynamics of $\braket{\hat{S}_x}$ in the Ising models in both 1D and 2D with minor deviations in 2D XYZ at long times (Fig.~\ref{NNResult}(a--c)). For $\Delta \hat{S}_x$, DTWA only captures the qualitative trend (Fig.~\ref{NNResult}(d--f)). ML-MCTDH exactly reproduces the dynamics across all cases for both $\braket{\hat{S}_x}$ and $\Delta \hat{S}_x$, particularly accounting for the small changes over very short time intervals while DTWA incorrectly smoothens such rapid fluctuations. 


\section{Conclusions and Outlook}
\label{Conclusion}
ML-MCTDH has been the workhorse for the numerical treatment of the high-dimensional dynamics of complex molecular systems with many degrees of freedom \cite{10.1063/1.1580111,10.1063/1.2363195,Wang2007-ds,doi:10.1021/jp072217m}. The flexibility offered by the ML-MCTDH in choosing the primitives, modifying the tree structure, and controllably optimizing the truncation makes it possible to tailor for the physical system of interest. This has led to adapting the ML-MCDTDH framework for many-body systems of bosons \cite{PhysRevLett.99.030402,PhysRevA.77.033613,10.1063/1.3173823,10.1063/1.4975662,10.1063/1.5140984} and fermions \cite{JürgenZanghellini_2004,PhysRevA.71.012712,10.1063/5.0028116}. Recently, it has also been utilized to investigate the ground-state properties of disordered spin models \cite{PhysRevResearch.5.023135}{ and to study the dynamics of centrally coupled systems with many-body interacting baths \cite{PhysRevB.103.134201}. Expanding on these previous research, the present study employs the ML-MCTDH approach to simulate many-body spin dynamics. }

In this study, we investigated different limits of the Heisenberg model across various settings. The key implication of our results is that ML-MCTDH provides an ideal framework for the long-time dynamics of on-site observables and more importantly two-point correlations, which are typically hard to obtain with existing numerical recipes for generic spin models \cite{Calabrese_2005,Schuch_2008,PhysRevLett.93.040502,PhysRevLett.97.157202,PhysRevLett.100.030504}. In the Ising limit, ML-MCTDH becomes exact for the dynamics of spins with power-law interactions with $\alpha=3,6$ including the $\alpha=0$ edge case with all-to-all couplings. This is evidenced by the perfect overlaps of the time-evolution profiles of collective spin observables with the analytical results in both 1D and 2D. Perfect agreement with the exact analytical calculations persists in the case of random disorder in the Ising limit for $\alpha=3,6$ albeit with minor deviations at short times for the all-to-all case. In the XYZ limit, benchmarks with exact diagonalization (ED) show that ML-MCTDH captures the dynamics exactly in 1D but with minor deviations in 2D. Comparative analysis with the DTWA indicates that ML-MCTDH is a better-suited method for gaining insights into long-time correlation dynamics across all simulation cases. In particular, for the XYZ limit, ML-MCTDH performs better in capturing both the on-site observable and correlation dynamics across all the simulation scenarios. {Our results based on entanglement dynamics and convergence analyses show that the rate of entanglement growth strongly depends on the interaction range and the presence of disorder. In disorder-free cases as well as disordered cases with $\alpha=3,6$, a limited number of SPFs are sufficient to accurately capture the dynamics. For the disordered model with all-to-all interactions, the entanglement grows rapidly at early times. This has implications in the classical simulability of disordered models with all-to-all connectivity with ML-MCTDH due to the unfavorable scaling of the number of required SPFs for accurate results. }

Our findings demonstrate that the ML-MCTDH method can accurately capture the dynamics of the Heisenberg model in various settings, as evidenced by benchmarking against exact methods and performing a comparative analysis with the DTWA. Therefore, a natural next step would be to use it as tool to study intriguing non-equilibrium phenomenons like quench dynamics \cite{Essler_2016,PhysRevX.8.021069}, thermalization \cite{RevModPhys.83.863,Gogolin_2016}, dynamical quantum phase transitions \cite{Heyl_2018,10.1063/1.4969869,PhysRevB.104.115133}, and dynamics of correlation spreading \cite{Lieb1972,PhysRevA.108.023301}. {As mentioned above, the rapid entanglement growth observed in all-to-all interacting disordered models significantly challenges classical simulability, even for short-time dynamics, necessitating a dedicated analysis of SPF scaling and the tree structure for optimal state representation}. Moreover, the scope in wavepacket dynamics includes optimal selection of trees, making the method more competitive for future applications \cite{10.1063/5.0035581,10.1063/1.5130390}. Recently, it has been shown that the ML-MCTDH framework shares similarities with the tree-tensor network architecture in certain aspects \cite{PhysRevA.74.022320,doi:10.1080/00268976.2024.2306881,10.1063/5.0218773}. Our findings may motivate the exploration of established ideas in ML-MCTDH to tree-tensor network framework and vice versa. 

\begin{acknowledgments}
{We are grateful to the anonymous referees, whose suggestions and insights prompted us to investigate entanglement dynamics and convergence analyses, thereby making our work more complete in characterizing the ML-MCTDH method.} This work is funded by the Cluster of Excellence
		``CUI: Advanced Imaging of Matter'' of the Deutsche Forschungsgemeinschaft (DFG) - EXC 2056 - Project ID 390715994. This work is funded by the German Federal Ministry of Education and Research within the funding program ``quantum technologies - from basic research to market" under contract 13N16138.
\end{acknowledgments}

\appendix
\renewcommand\thefigure{\thesection.\arabic{figure}} 
\section{ML-MCTDH}
  
\setcounter{figure}{0}    

\label{MLMCTDH}

\begin{figure*}
 \centering
  \includegraphics[width=180mm]{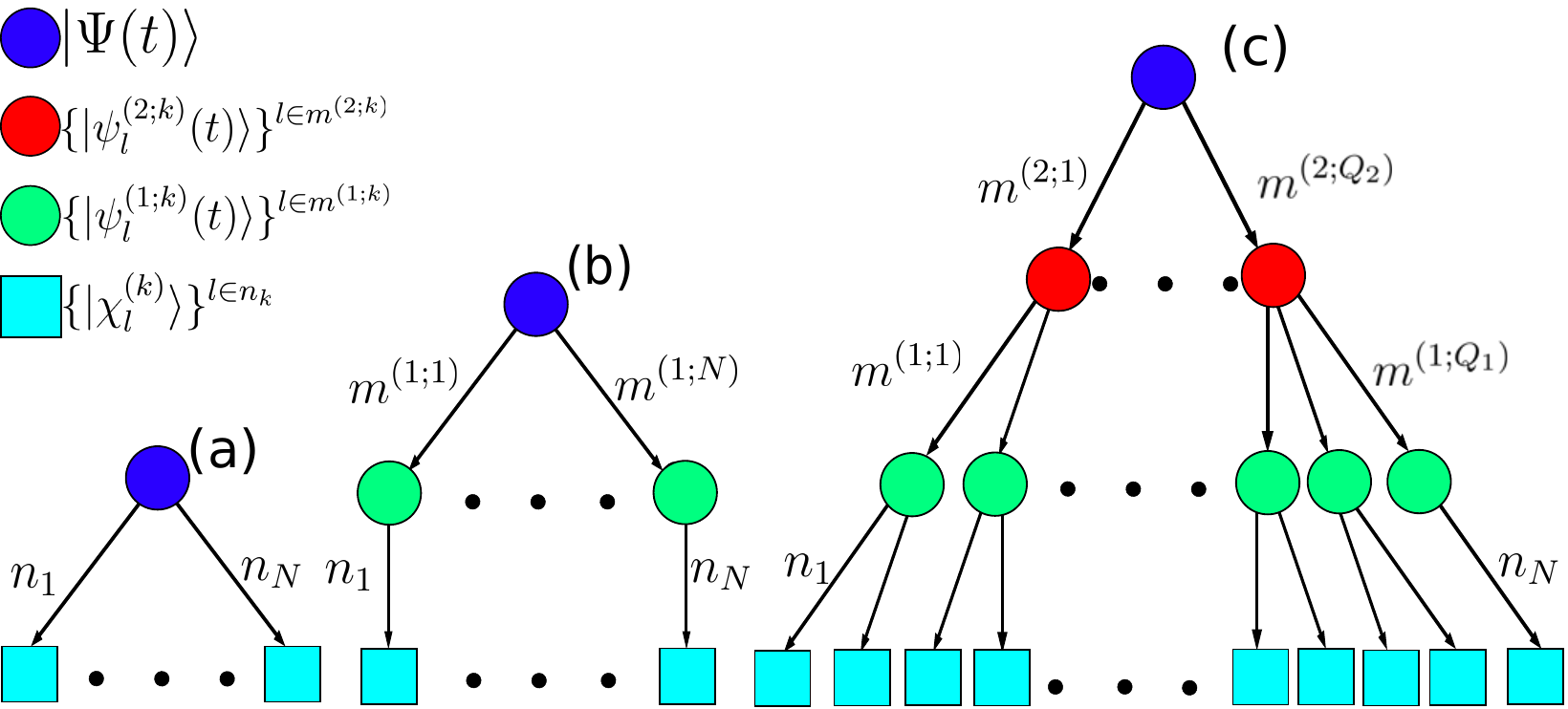}
\caption{Diagrammatic representation of the standard (a), MCTDH (b) and a two-layer ML-MCTDH (c) ansatz for the many body wavefunction $\ket{\Psi(t)}$ of a system with $N$ degrees of freedom. 
The cyan boxes represent the time-independent basis states, while the green and red circles correspond to the time-dependent basis states of layers 1 and 2, respectively. The blue circle denotes the many-body wavefunction.}
\label{TreeStructure}
\end{figure*}

The standard approach for solving the time-dependent Schrödinger equation involves propagating a wavepacket, expressed in terms of time-independent basis states with time-dependent expansion coefficients. This is illustrated for a system with $N$ physical degrees of freedom in Fig.~\ref{TreeStructure}(a).
The top node $\ket{\Psi(t)}$ (blue circle) is expressed as
\begin{equation}
    \ket{\Psi(t)} = \sum_{j_1=1}^{n_1} \cdots \sum_{j_N=1}^{n_N} A_{j_1 \cdots j_N}(t) \bigotimes_{k=1}^{N} \ket{ \chi_{j_k}^{(k)}} ,
    \label{StandardAnsatz}
\end{equation}
where $A_{j_1 \cdots j_N}$(t) denote the time-dependent expansion coefficients, $n_k$ represent the number of internal states for  $k$th degree of freedom, and the $\ket{\chi_{j_k}^{(k)}}$ (cyan boxes) are the time-independent basis functions for $k$th degree of freedom. \\
The time evolution of the many-body wave function $ \ket{\Psi(t)}$ is governed by the Dirac-Frenkel variational principle \cite{Dirac_1930},
\begin{equation}
    \braket{ \delta \Psi(t) | (i \partial_t - \hat{H}) | \Psi(t)} = 0.
    \label{Dirac_Frekel}
\end{equation}
By inserting the wavefunction ansatz (\ref{StandardAnsatz}) in Eq.~(\ref{Dirac_Frekel}), one obtains the equation of motion for the expansion coefficients $A_{j_1 \cdots j_N}(t)$,
\begin{equation}
\begin{split}
    i &\Dot{A}_{j_1  \cdots j_N}(t)   \\
    & = \sum_{l_1=1}^{n_1} \cdots \sum_{l_N=1}^{n_N} \bra{ \chi_{j_1}^{(1)} \cdots \chi_{j_N}^{(N)} } \hat{H} \ket{ \chi_{l_1}^{(1)} \cdots \chi_{l_N}^{(N)}} A_{l_1 \cdots l_N}(t),
    \end{split}
\end{equation}
which can be solved numerically using standard time integration methods. For a spin-$1/2$ system, $N$ refers to the number of sites, and $n_k=2$ with $\ket{\chi_{1}^{(k)}} = \ket{\uparrow}$ and $\ket{\chi_{2}^{(k)}} = \ket{\downarrow}$. This approach is limited to small system sizes as the number of coefficients  $\prod_{k=1}^N n_k$ scale exponentially with $N$ ($2^N$ for a spin-$1/2$ system).

MCTDH addresses this problem using a small set of time-dependent basis states for each individual physical degree of freedom. As shown in Fig.~\ref{TreeStructure}(b), the top node $\ket{\Psi(t)}$ is represented as \begin{equation}
        \ket{\Psi(t)} = \sum_{j_1=1}^{m^{(1;1)}} \cdots \sum_{j_N=1}^{m^{(1;N)}} A^{(1)}_{j_1 \cdots j_N}(t) \bigotimes_{k=1}^{N} \ket{ \psi_{j_k}^{(1;k)} (t)},
        \label{MCTDH_ansatz}
\end{equation}
where $A^{(1)}_{j_1 \cdots j_N}$ denotes the MCTDH expansion coefficients while the $\ket{\psi_{j_k}^{(1;k)}(t)}$ (green circle) denotes the $j_{k}$th time-dependent basis function for the $k$th degree of freedom which are referred to as single particle functions (SPFs). These time-dependent basis functions are chosen such that they form an orthonormal basis set and remain orthonormal throughout the time evolution. The numbers $m^{(1;k)}$ specify how many such functions are used for the $k$th degree of freedom. Each SPF, in turn, is represented with respect to the time-independent basis states of the standard ansatz (\ref{StandardAnsatz}),
\begin{equation}
    \ket{\psi_{j_k}^{(1;k)} (t)} = \sum_{l=1}^{n_k} c_{j_{k};l}^{(k)} (t) \ket{ \chi_{l}^{(k)}}. 
    \label{SPF_Primitive}
\end{equation}
The equations of motions for both the coefficients $A_{j_1 \cdots j_N}^{(1)}(t)$ and the SPFs $\ket{ \psi_{j_k}^{(1;k)} (t)}$ can be obtained by using the Dirac-Frenkel variational principle, the details can be found in \cite{BECK20001}. 

The MCTDH ansatz allows us to truncate the Hilbert space by choosing the optimal number of SPFs, $m^{(1;k)}$, for each degree of freedom. The Hilbert space dimensions of the top node $\ket{\Psi(t)}$ reduces from $\prod_{k=1}^N n_k$ in terms of time-independent basis states to $\prod_{k=1}^N m^{(1;k)}$ in terms of SPFs. Therefore, $m^{(1;k)}\ll n_k$ is chosen to achieve a significant reduction in the computational effort. For a spin-$1/2$ system, the specific case $m^{(1;k)}=1$ captures only mean-field effects. However, understanding the dynamics of strongly interacting spin systems necessitates a more accurate description incorporating beyond-mean-field correlations and entanglement, and therefore $m^{(1;k)} > 1$ applies.

\begin{figure}[]
\begin{center}
  \includegraphics[height=45mm]{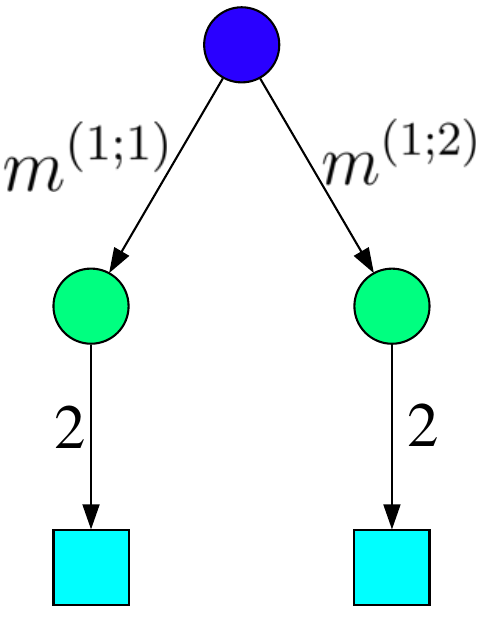} 
\end{center}
\caption{Diagrammatic representation of the MCTDH ansatz for a spin-$1/2$ system with 2 spins.} 
\label{MCTDH_eg}
\end{figure}

In the following, the MCTDH ansatz is explicitly shown for a simple case to demonstrate how it truncates the Hilbert space. Consider the case of a spin-$1/2$ lattice with 2 spins. Hence, $N=2$, $n_1=n_2=2$, i.e., $\ket{ \chi_{1}^{(k)}} = \ket{\uparrow}$ and $\ket{ \chi_{2}^{(k)}} = \ket{\downarrow}$. So, the standard ansatz according to Eq.~(\ref{StandardAnsatz}) is,
\begin{equation}
\begin{split}
    \ket{\Psi(t)} = A_{11}(t)\ket{\uparrow \uparrow} &+ 
    A_{12}(t)\ket{\uparrow \downarrow} + \\
    & A_{21}(t)\ket{\downarrow \uparrow} + A_{22}(t)\ket{\downarrow \downarrow}.
    \label{2spinStandard}
    \end{split}
\end{equation}
Above, note that the time-dependent expansion coefficients are independent of each other. For the MCTDH ansatz, shown in Fig.~\ref{MCTDH_eg}, we consider two cases $m^{(1;1)}=m^{(1;2)}=1$ and $m^{(1;1)}=m^{(1;2)}=2$.

For the case with $m^{(1;1)}=m^{(1;2)}=1$, the MCTDH ansatz is given by
\begin{equation}
    \ket{\Psi(t)} = A_{11}^{(1)}(t)\ket{\psi_1^{(1;1)}(t) \psi_1^{(1;2)}(t)}.
\end{equation}
The above equation can be expanded further in terms of time-independent basis functions by using Eq.~(\ref{SPF_Primitive}),
\begin{equation}
\begin{split}
    \ket{\Psi(t)} = A_{11}^{(1)}(t) &\left[   c_{1;\uparrow}^{(1)}(t)c_{1;\uparrow}^{(2)}(t)\ket{\uparrow \uparrow} +c_{1;\uparrow}^{(1)}(t)c_{1;\downarrow}^{(2)}(t)\ket{\uparrow \downarrow} + \right.\\
     & \left. c_{1;\downarrow}^{(1)}(t)c_{1;\uparrow}^{(2)}(t)\ket{\downarrow \uparrow} +
    c_{1;\downarrow}^{(1)}(t)c_{1;\downarrow}^{(2)}(t)\ket{\downarrow \downarrow} \right].
\end{split}
    \label{2spin1MCTDH}
\end{equation}
Unlike Eq.~\ref{2spinStandard}, the time-dependent expansion coefficients are not independent of each other in the above equation. This results in a truncation of the Hilbert space as we can no longer represent all the possible states. For example, it is not possible to represent a state of the form $\ket{\Psi(t)} = a(t)\ket{\uparrow\uparrow} + b(t) \ket{\downarrow \downarrow}$ since the expansion coefficients of $\ket{\uparrow \downarrow}$ and $\ket{\downarrow \uparrow}$ cannot vanish.

For the case with $m^{(1;1)}=m^{(1;2)}=2$, the MCTDH ansatz is given by
\begin{equation}
    \begin{split}
     \ket{\Psi(t)} = & A_{11}^{(1)}(t)\ket{\psi_1^{(1;1)}(t)\psi_1^{(1;2)}(t)} +\\ & A_{12}^{(1)}(t)\ket{\psi_1^{(1;1)}(t)\psi_2^{(1;2)}(t) } +  \\& A_{21}^{(1)}(t)\ket{\psi_2^{(1;1)}(t)\psi_1^{(1;2)}(t)} + \\
    &
    A_{22}^{(1)}(t)\ket{\psi_2^{(1;1)}(t)\psi_2^{(1;2)}(t)}.
\end{split}
 \label{2spin2MCTDH}
\end{equation}
The above equation can be expanded further in terms of time-independent basis functions,
\begin{equation}
\begin{split}
    \ket{\Psi(t)} = & \left[ A_{11}^{(1)}c_{1;\uparrow}^{(1)}(t)c_{1;\uparrow}^{(2)}(t) + A_{12}^{(1)}c_{1;\uparrow}^{(1)}(t)c_{2;\uparrow}^{(2)}(t) + \right.\\
    &
     \left. A_{21}^{(1)}c_{2;\uparrow}^{(1)}(t)c_{1;\uparrow}^{(2)}(t) + A_{22}^{(1)}c_{2;\uparrow}^{(1)}(t)c_{2;\uparrow}^{(2)}(t) \right] \ket{\uparrow\uparrow} + \\
    &\left[ A_{11}^{(1)}c_{1;\uparrow}^{(1)}(t)c_{1;\downarrow}^{(2)}(t) + A_{12}^{(1)}c_{1;\uparrow}^{(1)}(t)c_{2;\downarrow}^{(2)}(t) + \right. \\
    & \left. A_{21}^{(1)}c_{2;\uparrow}^{(1)}(t)c_{1;\downarrow}^{(2)}(t) + A_{22}^{(1)}c_{2;\uparrow}^{(1)}(t)c_{2;\downarrow}^{(2)}(t) \right] \ket{\uparrow \downarrow} + \\
    & \left[ A_{11}^{(1)}c_{1;\downarrow}^{(1)}(t)c_{1;\uparrow}^{(2)}(t) + A_{12}^{(1)}c_{1;\downarrow}^{(1)}(t)c_{2;\uparrow}^{(2)}(t) + \right. \\
    & \left. A_{21}^{(1)}c_{2;\downarrow}^{(1)}(t)c_{1;\uparrow}^{(2)}(t) + A_{22}^{(1)}c_{2;\downarrow}^{(1)}(t)c_{2;\uparrow}^{(2)}(t) \right] \ket{\downarrow\uparrow} + \\
    & \left[ A_{11}^{(1)}c_{1;\downarrow}^{(1)}(t)c_{1;\downarrow}^{(2)}(t) + A_{12}^{(1)}c_{1;\downarrow}^{(1)}(t)c_{2;\downarrow}^{(2)}(t) + \right. \\ 
    & \left. A_{21}^{(1)}c_{2;\downarrow}^{(1)}(t)c_{1;\downarrow}^{(2)}(t) + A_{22}^{(1)}c_{2;\downarrow}^{(1)}(t)c_{2;\downarrow}^{(2)}(t) \right] \ket{\downarrow\downarrow}.
\end{split}
\end{equation}
By introducing just one more SPF for each degree of freedom, we have made the MCTDH ansatz more complex and richer than Eq.~(\ref{2spin1MCTDH}) due to the introduction of more time-dependent coefficients. This reduces the dependency between time-dependent expansions coefficients of time-independent basis functions. As a result, the ansatz can represent more number of possible states. In this specific case, the MCTDH ansatz becomes exact as dimensions of the Hilbert space of the many-body wavefunction in Eq.~(\ref{2spin2MCTDH}) are the same as the dimensions of the Hilbert space of the many-body wavefunction in Eq.~(\ref{2spinStandard}).

ML-MCTDH improves upon MCTDH by forming a layered structure of time-dependent basis states to represent the many-body wavefunction \cite{10.1063/1.1580111}. In this approach, the physical degrees of freedom are grouped into a single node (green circle), which is represented by a small set of SPFs, forming the first layer of ML-MCTDH ansatz as illustrated in Fig.~\ref{TreeStructure}(c). This contrasts with MCTDH ansatz depicted in Fig.~\ref{TreeStructure}(b) where a single node corresponds to a single physical degree of freedom. ML-MCTDH also allows additional layers, each introducing new nodes that group degrees of freedom from the layer below, recursively creating a tree structure. In Fig.~\ref{TreeStructure}(c), a two-layer ML-MCTDH ansatz is illustrated, where the top node $\ket{\Psi(t)}$ is expressed as
\begin{equation}
    \ket{\Psi(t)}=\sum_{j_1=1}^{m^{(2;1)}} \cdots \sum_{j_{Q_2=1}}^{m^{(2;Q_2)}} A^{(2)}_{j_1 \cdots j_{Q_{2}}}(t) \bigotimes_{k=1}^{Q_{2}} \ket{ \psi_{j_k}^{(2;k)} (t)},
    \label{L2Ansatz}
\end{equation}
where $Q_{2}$ specifies the number of layer 2 (L2) degrees of freedom, $A^{(2)}_{j_1 \cdots j_{Q_{2}}}(t)$ denotes the expansion coefficients and $\ket{ \psi_{j_k}^{(2;k)} (t)}$ (red circle) denotes the $j_k$th SPF for the $k$th L2 degree of freedom. The number $m^{(2;k)}$ specify the number of SPFs for the 
$j_k$th L2 degree of freedom. These L2 SPFs in turn are represented with respect to the time-dependent basis set of layer 1 (L1) with time-dependent expansion coefficients,
\begin{equation}
\begin{split}
    \ket{ \psi_{j_k}^{(2;k)} (t)}=&\sum_{j_{\alpha_2}=1}^{m^{(1;{\alpha_2})}} \cdots \sum_{j_{\beta_2}=1}^{m^{(1;\beta_2)}} A^{(1;k)}_{j_k;j_{\alpha_2} \cdots j_{\beta_2}}(t) \\
    &\bigotimes_{i={\alpha_2}}^{{\beta_2}} \ket{ \psi_{j_i}^{(1;i)} (t)}.
\end{split}
\end{equation}

Here, $\alpha_2 = \alpha_2(k)$ and $\beta_2 =\beta_2(k)$ refer to the index of the first and last element of the unit belonging to the $k$th L2 degree of freedom such that $\alpha_2(1)=1$ and $\beta_2(Q_2)=Q_1$, where $Q_1$ specifies the number of layer 1 (L1) degrees of freedom. Similar to Eq.~(\ref{SPF_Primitive}), these L1 SPFs can be further expressed in terms of the time-independent basis states with time-dependent expansion coefficients,
\begin{equation}
\begin{split}
        \ket{\psi^{(1;i)}_{j_i}(t)}=&\sum_{j_{\alpha_1}=1}^{n_{{\alpha_1}}} \cdots \sum_{j_{\beta_1}=1}^{n_{\beta_1}} c^{(i)}_{j_k;j_i;j_{\alpha_1} \cdots j_{\beta_1}}(t) \\
    &\bigotimes_{l={\alpha_1}}^{{\beta_1}} \ket{ \chi_{j_l}^{(l)} (t)}.
\end{split}
\end{equation}
Above, $\alpha_1 = \alpha_1(i)$ and $\beta_1 =\beta_1(i)$ refer to the index of the first and last element of the unit belonging to the $i$th L1 degree of freedom such that $\alpha_1(1)=1$ and $\beta_1(Q_1)=N$. 

The Hilbert space dimension of the top node $\ket{\Psi(t)}$ reduces from $\prod_{k=1}^N n_k$ in terms of time-independent basis states to $\prod_{k=1}^{Q_2} m^{(2;k)}$ in terms of L2 SPFs. Hence, ML-MCTDH  allows us to choose the optimal numbers of SPFs and create a tree structure specifically tailored to the physical problem under consideration. 
The equations of motion for the expansion coefficients and SPFs are derived using the Dirac-Frenkel variational principle, as detailed in \cite{10.1063/1.1580111}.
\begin{figure}[]
\begin{center}
  \includegraphics[height=45mm]{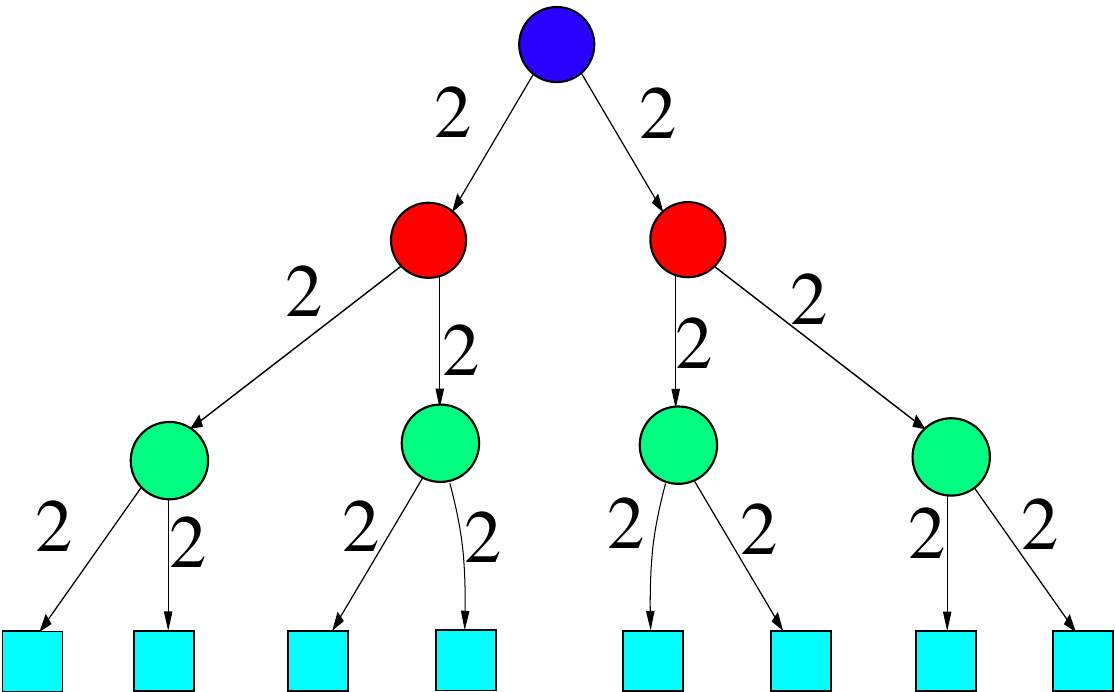} 
\end{center}
\caption{Diagrammatic representation of the binary tree ansatz  employed for a spin-$1/2$ system with 8 spins. For this ansatz, we choose $m^{(2;k)}=2$ and $m^{(1;k)}=2$.} 
\label{ML-MCTDH_eg}
\end{figure}

To explicitly illustrate the ML-MCTDH ansatz, we consider an example of a spin-$1/2$ system with 8 spins. Fig.~\ref{ML-MCTDH_eg} depicts the binary tree used in this ansatz, where $Q_2=2$ and $Q_1=4$. We set $m^{(2;i)}=2$ and $m^{(1;i)}=2$. Following Eq.(\ref{L2Ansatz}), the top node is expanded using the L2 SPFs as
    \begin{equation}
    \begin{split}
     \ket{\Psi(t)} = & A_{11}^{(2)}(t)\ket{\psi_1^{(2;1)}(t)\psi_1^{(2;2)}(t)} +\\ & A_{12}^{(2)}(t)\ket{\psi_1^{(2;1)}(t)\psi_2^{(2;2)}(t)} +  \\& A_{21}^{(2)}(t)\ket{\psi_2^{(2;1)}(t)\psi_1^{(2;2)}(t) } + \\
    &
    A_{22}^{(2)}(t)\ket{\psi_2^{(2;1)}(t)\psi_2^{(2;2)}(t)}.
\end{split}
 \label{8spinL2MLCTDH}
\end{equation}
Rather than expanding the top node explicitly in terms of time-independent basis states, we expand the SPFs of the first node in each layer into the basis states of the layer below. Thus, the L2 SPFs of the first node are expressed in terms of the L1 SPFs as follows:
\begin{equation}
    \begin{split}
     \ket{\psi_1^{(2;1)}(t)} = & A_{1;11}^{(1;1)}(t)\ket{\psi_1^{(1;1)}(t)\psi_1^{(1;2)}(t)} +\\ & A_{1;12}^{(1;1)}(t)\ket{\psi_1^{(1;1)}(t)\psi_2^{(1;2)}(t) } +  \\& A_{1;21}^{(1;1)}(t)\ket{\psi_2^{(1;1)}(t)\psi_1^{(1;2)}(t)} + \\
    &
    A_{1;22}^{(1;1)}(t)\ket{\psi_2^{(1;1)}(t)\psi_2^{(1;2)}(t)}.
\end{split}
 \label{8spinL1MLCTDH}
\end{equation}
The L1 SPFs of the first node can be further expanded in terms of time-independent basis states, given by
\begin{equation}
    \begin{split}
     \ket{\psi_1^{(1;1)}(t)} = & c^{(1)}_{1;1;\uparrow\uparrow}(t)\ket{\uparrow\uparrow} +\\ & c^{(1)}_{1;1;\uparrow\downarrow}(t)\ket{\uparrow\downarrow} +  \\& c^{(1)}_{1;1;\downarrow\uparrow}(t)\ket{\downarrow\uparrow} + \\
    &
    c_{1;1;\downarrow\downarrow}^{(1)}(t)\ket{\downarrow\downarrow}.
\end{split}
 \label{8spinMLCTDH}
\end{equation}
 For this tree structure, $\alpha_2(k)=\alpha_1(k) = 2k-1$ and $\beta_2(k)=\beta_1(k)=2k$. Through this ansatz, the Hilbert space dimensions reduce from $2^{8}$ in terms of time-independent basis states to $2^{2}$ in terms of L2 SPFs.

Now, we discuss the tree structures employed throughout this work. All the tree structures are symmetric. We use a simple notation to represent the tree structure. For instance, the tree in Fig.~\ref{ML-MCTDH_eg} is denoted by $8\xrightarrow[]{n_k \equiv2}4\xrightarrow[]{m^{(1;k)}\equiv2}2\xrightarrow[]{m^{(2;k)}\equiv2}1$. 

In the Ising limit for 1D chains with long-range interactions (where $L=32$), we utilize  $32\xrightarrow[]{2}16\xrightarrow[]{4}4\xrightarrow[]{12}1$. For systems with nearest-neighbour interaction (where $L=128$), we employ $128\xrightarrow[]{2}64\xrightarrow[]{4}16\xrightarrow[]{10}4\xrightarrow[]{18}1$. 
For the disordered case (where $L=32$), we utilized $32\xrightarrow[]{2}16\xrightarrow[]{4}4\xrightarrow[]{16}1$ for $\alpha=3,6$ and $32\xrightarrow[]{2}16\xrightarrow[]{4}4\xrightarrow[]{22}1$ for $\alpha=0$.

In the XYZ limit for 1D chains (where $L=16$), we employ $16\xrightarrow[]{2}4\xrightarrow[]{10}1$ for $\alpha=0$ and $16\xrightarrow[]{2}4\xrightarrow[]{12}1$ in the cases of $\alpha =3,6$.
\begin{figure}[]
\begin{center}
  \includegraphics[width=85mm]{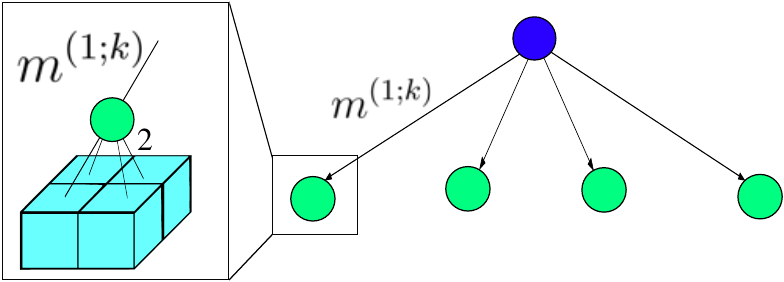} 
\end{center}
\caption{Diagrammatic representation of the tree structure employed for 2D lattice ($L=4\times 4$).} 
\label{2D_lattice}
\end{figure}

For 2D lattices, we employ $16\xrightarrow[]{2}4\xrightarrow[]{m^{(1;k)}}1$ as depicted in Fig.~\ref{2D_lattice}. In the Ising limit, we assign $m^{(1;k)}=8$, while in the XYZ limit, we set $m^{(1;k)}=14$.

\section{DTWA}
\label{DTWA}
The discrete truncated Wigner approximation is a semiclassical method based on the sampling of a discrete Wigner function \cite{PhysRevX.5.011022}. The time-evolution is carried out via classical trajectories. The quantum uncertainty in the initial state is accounted for by averaging over different initial conditions determined by the discrete Wigner function. To apply this method, we have to compute the classical equations of motion for the spin components of each spin $i$: $s_i^x, s_i^y, s_i^z$. These can be obtained from the classical Hamiltonian function, corresponding to the Hamiltonian operator in Eq.~(\ref{XYZHamiltonian}), given by
\begin{equation}
        H_C = - \sum_{\substack{i,j=1\\i<j}}^{L} \left[ J_{ij}^{x} {s}_{i}^x {s}_{j}^x + J_{ij}^{y} {s}_{i}^y {s}_{j}^y + J_{ij}^{z} {s}_{i}^z {s}_{j}^z \right],
\end{equation}
via
\begin{equation}
    \Dot{s}_i^\mu = \{ s_i^\mu, H_C\} = 2 \sum_\beta \epsilon_{\alpha \beta \gamma} s_i^\gamma \frac{\partial H_C}{\partial s_i^\beta},
\end{equation}
where $\{.,.\}$ denote the Poisson bracket and $\epsilon$ the fully antisymmetric tensor.

After obtaining the equations of motion, we numerically integrate them by choosing a large number $n_t$ of different random initial configurations corresponding to our initial state. As stated above, we perform our calculation for an initial state where all spins are aligned along the x-axis, $\ket{\rightarrow \rightarrow \cdots \rightarrow}$. This corresponds to choosing $s_i^x(0)=1$ and randomly choosing $s_i^y(0),s_i^z(0) =\pm1$ for each spin $i$. The final expectation values of observables are calculated by averaging the results for the corresponding observable over all initial conditions. In all our calculations, we set $n_t=10000$.

For example, for the Ising interaction ($J_{ij}^{x,y}=0$), the classical equations for the spin components are given by
\begin{equation}
 \Dot{s}_i^x = -2 s_i^y \sum_j J_{ij}^z s_j^z, 
\end{equation}
\begin{equation}
 \Dot{s}_i^y = -2 s_i^x \sum_j J_{ij}^z s_j^z,  
\end{equation}
and,
\begin{equation}
 \Dot{s}_i^z = 0. 
\end{equation}
Solving these equations yields
\begin{equation}
s_i^{\pm}(t) = s_i^{\pm}(0) \exp(\pm 2it\sum_j J_{ij}^zs_j^z ),
\label{Onepoint_dtwa}
\end{equation}
where $s_j^{\pm}=(s_j^x \pm i s_j^y)/2$. Approximating the above sum via a random sampling of $s_i^z$ taking the values of +1,-1 results in exact time evolution \cite{Worm_2013,PhysRev.107.46} for a local operator,
\begin{equation}
         \braket{\hat{\sigma}_i^{\pm}}(t)_{DTWA}=  \braket{\hat{\sigma}_i^{\pm}}(0) \prod_{k\ne i} \cos(2t J^z_{k,i}) = \braket{\hat{\sigma}_i^{\pm}} (t)_{exact} .
         \label{one-point}
\end{equation}
The same calculation is possible for two-point correlation functions,
\begin{equation}
\begin{split}
    \braket{\hat{\sigma}_i^{\pm}\hat{\sigma}_j^{\pm}}(t)_{DTWA} &=  \braket{ \hat{\sigma}_i^{\pm} \hat{\sigma}_j^{\pm}} (0) \prod_{k} \cos \left[ 2t (J^z_{k,i} + J^z_{k,j}) \right] \\
    & = \braket{\hat{\sigma}_i^{\pm}\hat{\sigma}_j^{\pm}}(t)_{exact} \cos^2 (2tJ_{ij}^z),
    \end{split}
    \label{two-point1}
\end{equation}
and
\begin{equation}
\begin{split}
    \braket{\hat{\sigma}_i^{\mp}\hat{\sigma}_j^{\pm}}(t)_{DTWA} &=  \braket{ \hat{\sigma}_i^{\mp} \hat{\sigma}_j^{\pm}} (0) \prod_{k} \cos \left[ 2t (J^z_{k,i} - J^z_{k,j}) \right] \\
    & = \braket{\hat{\sigma}_i^{\mp}\hat{\sigma}_j^{\pm}}(t)_{exact} \cos^2 (2tJ_{ij}^z).
    \end{split}
    \label{two-point2}
\end{equation}
The Eqs.~(\ref{one-point}-\ref{two-point2}) illustrate the reason for the agreement in $\braket{\hat{S}_x}$ and the discrepancies in $\Delta\hat{S}_x$ for the Ising interaction discussed in Sec.~\ref{Results}.
\bibliography{references} 

\end{document}